\documentclass[10pt,conference]{IEEEtran}
\IEEEoverridecommandlockouts

\usepackage{color}
\usepackage{amsmath, bm, amssymb}
\usepackage{xfrac}
\usepackage{amsfonts}
\usepackage{enumitem}
\usepackage{tabularx}
\usepackage[T1]{fontenc}
\usepackage{listings}
\usepackage{xspace} 
\usepackage{subfigure}
\usepackage{tcolorbox}
\usepackage{graphicx}
\usepackage{textcomp}
\usepackage{pifont}
\usepackage{booktabs}
\usepackage{multirow}
\usepackage{adjustbox}
\usepackage{makecell}
\usepackage[referable]{threeparttablex}
\usepackage{balance}
\usepackage{cite}

\newcolumntype{Y}{>{\centering\arraybackslash}X}

\newcommand{\app}{\textsc{InferROI}\xspace}

\newcommand{\gptleak}{\textsc{GPTLeak}\xspace}
\newcommand{\gptleakexp}{\textsc{GPTLeak}-\textit{exp}\xspace}
\newcommand{\gptleakroi}{\textsc{GPTLeak}-\textit{roi}\xspace}

\newcommand{\Comment}[1]{}

\newcommand{\gptfour}{GPT-4\xspace}

\usepackage{algorithm,algpseudocode}
\algrenewcommand\algorithmicindent{1.0em}
\algnewcommand{\LeftComment}[1]{\(\triangleright\) #1}

\usepackage{stackengine}

\usepackage{listings}
\definecolor{codegreen}{rgb}{0,0.6,0}
\definecolor{codegray}{rgb}{0.5,0.5,0.5}
\definecolor{codepurple}{rgb}{0.58,0,0.82}
\newcounter{finding}
\newcommand{\finding}[1]{\refstepcounter{finding}
	\begin{center}
		\begin{tcolorbox}[colback=gray!10,colframe=black!50,width=1\columnwidth,arc=0.5mm, auto outer arc,boxrule=0.5pt,boxsep=3pt,left=1pt,right=1pt,top=0pt,bottom=0pt]
		\textbf{Summary:} #1
		\end{tcolorbox}
	\end{center}
}

\usepackage[T1]{fontenc}
\usepackage[scaled]{sourcecodepro}
\usepackage{microtype}
\newcommand{\inlinecode}[1]{%
	{\texttt{\adjustcode\footnotesize #1\relax}}%
}
\newcommand{\adjustcode}[1]{%
  \ifx#1\relax
    \else
    \ifx#1.%
      \kern-0.15em
      #1%
      \kern-0.15em
    \else
      #1%
    \fi
    \expandafter\adjustcode
  \fi
}

\newlength{\ColorBoxDepthReference}
\newlength{\ColorBoxHeightReference}
\newlength{\Width}%
\newcommand{\MyColorBox}[2][red]%
{%
	\settowidth{\Width}{#2}%
	\colorbox{#1}%
	{%      
		\raisebox{-\ColorBoxDepthReference}%
		{%
			\parbox[b][\ColorBoxHeightReference+\ColorBoxDepthReference][c]{\Width}{\centering#2}%
		}%
	}%
}
\newcommand\acqbox[1]{\MyColorBox[red!20]{#1}}
\newcommand\relbox[1]{\MyColorBox[green!20]{#1}}
\newcommand\validbox[1]{\MyColorBox[cyan!20]{#1}}

\newcommand\revise[1]{\textcolor{black}{#1}}
\newcommand\cameraready[1]{\textcolor{black}{#1}}

\begin{document}
% \title{Inferring Resource-Oriented Intentions using LLMs for Static Resource Leak Detection}
\title{Boosting Static Resource Leak Detection via LLM-based Resource-Oriented Intention Inference}

\author{
    \IEEEauthorblockN{Chong Wang\IEEEauthorrefmark{2}, Jianan Liu\IEEEauthorrefmark{2}, Xin Peng\IEEEauthorrefmark{2}, Yang Liu\IEEEauthorrefmark{3}, and Yiling Lou\IEEEauthorrefmark{2}\IEEEauthorrefmark{1}}
    \IEEEauthorblockA{
        \IEEEauthorrefmark{2}\textit{School of Computer Science and Shanghai Key Laboratory of Data Science, Fudan University, China}\\ 
        \IEEEauthorrefmark{3}\textit{School of Computer Science and Engineering, Nanyang Technological University, Singapore}\\
        \{wangchong20, pengxin@fudan.edu.cn, yilinglou@fudan.edu.cn\}@fudan.edu.cn\\
        yangliu@ntu.edu.sg
    }
    \thanks{\IEEEauthorrefmark{1} Yiling Lou is the corresponding author}
}

\maketitle

\begin{abstract}
Resource leaks, caused by resources not being released after acquisition, often lead to performance issues and system crashes. Existing static detection techniques rely on mechanical matching of predefined resource acquisition/release APIs and null-checking conditions to find unreleased resources, suffering from both (1) false negatives caused by the incompleteness of predefined resource acquisition/release APIs and (2) false positives caused by the incompleteness of resource reachability validation identification. 
To overcome these challenges, we propose \app, a novel approach that leverages the exceptional code comprehension capability of large language models (LLMs) to directly infer resource-oriented intentions (acquisition, release, and reachability validation) in code. \app first prompts the LLM to infer involved intentions for a given code snippet, and then incorporates a two-stage static analysis approach to check control-flow paths for resource leak detection based on the inferred intentions. 

We evaluate the effectiveness of \app in both resource-oriented intention inference and resource leak detection. Experimental results on the DroidLeaks and JLeaks datasets demonstrate \app achieves promising bug detection rate (59.3\% and 62.5\%) and false alarm rate (18.6\% and 19.5\%). Compared to three industrial static detectors, \app detects 14\textasciitilde 45 and 149\textasciitilde 485 more bugs in DroidLeaks and JLeaks, respectively. When applied to real-world open-source projects, \app identifies 29 unknown resource leak bugs (\revise{verified by authors}), with 7 of them being confirmed by developers. In addition, the results of an ablation study underscores the importance of combining LLM-based inference with static analysis. Finally, manual annotation indicated that \app achieved a precision of 74.6\% and a recall of 81.8\% in intention inference, covering more than 60\% resource types involved in the datasets.
\end{abstract}

\maketitle

\section{Introduction}

Resource leaks, stemming from the failure to release acquired resources (e.g., unclosed file handles), represent critical software defects that can lead to runtime exceptions or program crashes. Such leaks are pervasive in software projects~\cite{ghanavati20leak} and even appear in online code examples, even for some accepted answers in Stack Overflow posts~\cite{zhang2018code}.

To address this issue, researchers have proposed various automated techniques~\cite{torlak2010effective, wu2016light,kellogg2021lightweight,spotbugs,infer} for resource leak detection based on static analysis. These techniques primarily rely on two essential components. The first step involves identifying pairs of resource acquisition APIs and their corresponding resource release APIs;  Subsequently, the code is analyzed to verify whether the release API is not appropriately invoked after the acquisition API on the control-flow paths, before the acquired resources become unreachable.
For instance, consider the API pair for managing locks: \inlinecode{LockManager.acquireLock()} represents the resource acquisition API, and \inlinecode{LockManager.releaseLock()} denotes the corresponding resource release API. A resource leak occurs when \inlinecode{releaseLock()} is not subsequently called after \inlinecode{acquireLock()}.

%Existing techniques focus on analyzing while few focus on collecting RARs 
However, these existing techniques mainly rely on mechanical matching of predefined resource acquisition/release APIs and null-checking conditions to find unreleased resources, leading to the following two limitations. 
\textbf{\textit{(1) False negatives from the incompleteness of predefined resource acquisition/release APIs:}} the effectiveness of current resource leak detection techniques~\cite{torlak2010effective, wu2016light, kellogg2021lightweight} hinges on the completeness of predefined API pairs, since they cannot detect resource leaks related to un-predefined resource acquisition/release API pairs. For example, the widely adopted detectors (e.g., SpotBugs, Infer, and CodeInspection) are confined to detecting only a small subset of common resource classes in JDK and Android~\cite{liu2019droidleaks} by default. Thus, the resource leaks involving emerging or less common resources (e.g., \inlinecode{AndroidHttpClient} in \textit{apache httpclient} library) cannot be detected by these tools if the specific acquisition/release APIs for \inlinecode{AndroidHttpClient} are not additionally configured into these tools. 
\textbf{\textit{(2) False positives from the incompleteness of resource reachability validation identification:}} the accuracy of reachability analysis for resources along control-flow paths significantly influences the occurrence of false alarms in existing detection methods, since unreachable resources would not cause leaks even without being released after acquisition. Approaches like those introduced by Torlak et al.~\cite{torlak2010effective} determine resource reachability by scrutinizing \inlinecode{null}-checking conditions (e.g., \inlinecode{cur != null}) within \inlinecode{if}-statements. However, this approach falls short for certain resources with alternative means of determining reachability, such as through API calls (e.g., \inlinecode{!bank.isDisabled()}). Consequently, false alarms are triggered for resources that are actually \textit{unreachable} along specific paths.

In this paper, we propose \textbf{\textit{\app}}, a novel approach that leverages large language models (LLMs) to boost static resource leak detection. \textit{Our intuition is that LLMs have been trained on massive code and documentation corpus and thus have exceptional code comprehension capability~\cite{openai2023gpt4,bubeck2023sparks,petroni2019language,huang2022prompt}, suggesting their potential for inferring resource-oriented intentions \revise{(i.e., the code statements for resource acquisition/release and resource reachability validation)} in any given code snippet.} 
\app consists of two stages, i.e., resource-oriented intention inference and lightweight resource leak detection. (1) For a given code snippet, \app first prompts the LLM to directly infer the involved resource acquisition, release, and reachability validation intentions, without relying on any prior knowledge of resource API pairs. The intentions generated by LLMs (which are described in natural language) are then converted into \revise{structured} expressions, so as to facilitate the subsequent static analysis stage. (2) \app then incorporates a two-stage static analysis approach to check control-flow paths for resource leaks based on the inferred intentions. In particular, \app first identifies potential leak-risky paths based on the inferred resource acquisition/release APIs and then prunes leak-risky paths with unreachable resources based on the inferred reachability validation, so as to reduce false alarms.

%Rather than relying solely on mechanical matching of predefined resource acquisition/release APIs and \inlinecode{null}-checking conditions, our approach addresses the aforementioned challenges by directly inferring resource-oriented intentions (including resource acquisition, release, and reachability validation) combining resource management knowledge and code context understanding. In the approach, LLMs function as extensive knowledge bases with exceptional code understanding capabilities~\cite{openai2023gpt4,bubeck2023sparks,petroni2019language,huang2022prompt}, rendering them well-suited for inferring resource-oriented intentions. Upon receiving a code snippet, \app employs a prompt template to instruct the LLM in inferring the involved resource acquisition, release, and reachability validation intentions, eliminating the need for prior knowledge of resource API pairs. The LLM generates outputs in natural language, which are subsequently parsed into formal expressions of intentions. By aggregating these inferred intentions, \app proceeds to implement a lightweight static-analysis based algorithm to analyzes the control-flow paths extracted from the code, allowing for the detection of resource leaks. The algorithm takes into consideration the resource acquisition and release along individual control flow paths to find potential leaks. Then, it accounts for the impact of resource reachability across these paths on sibling branches, mitigating false alarms and enhancing the precision of leak detection. 

%our evalution 
While our approach is general and not limited to specific programming languages, we currently implement \app for Java given its prevalence. 
We conduct an extensive evaluation to assess the efficacy of \app in detecting resource leaks.
Firstly, we apply \app to 86 and 784 bugs sourced from the DroidLeaks~\cite{liu2019droidleaks} and JLeaks~\cite{icse/Jleaks} datasets, respectively, to evaluate its effectiveness in resource leak detection. The results reveal \app demonstrates a bug detection rate of 59.3\% with a false alarm rate of 18.6\% for DroidLeaks, and a bug detection rate of 62.5\% with a false alarm rate of 19.5\% for JLeaks. Compared to three industrial static detectors, \app detects 14\textasciitilde 45 and 149\textasciitilde 485 more bugs in DroidLeaks and JLeaks, respectively, while maintaining a comparable false alarm rate.
Secondly, we employ \app for resource detection in real-world open-source projects. \app uncovers 29 previously unknown resource leaks from 200 methods, 7 of which are confirmed by developers as of the submission time. In addition, an ablation study underscores the importance of combining LLM-based inference with static analysis. \revise{The generalizability of \app with different LLMs is confirmed by an evaluation involving Llmam3-8B and Gemma2-9B.}
Finally, manual annotation indicates that \app achieves a precision of 74.6\% and a recall of 81.8\% in intention inference, covering 19 (67.9\%) and 221 (60.1\%) resource types involved in DroidLeaks and JLeaks, respectively.

% summary for our contribution 
To summarize, this paper makes the following contributions:
\begin{itemize}[leftmargin=15pt]
    \item \textbf{A novel LLM-based perspective} that infers resource-oriented intentions (resource acquisition, release, and reachability validation) in code instead of mechanically matching predefined APIs. The inferred intentions can be applied to boost static resource leak detection techniques.
    
    \item \textbf{A lightweight detection approach \app} that combines the LLM-based intention inference and a static two-stage path analysis to detect resource leaks in code.
    
    \item \textbf{An extensive evaluation} that demonstrates the effectiveness of \app in inferring resource-oriented intentions and detecting diverse resource leaks in both existing datasets and real-world open-source projects. 
    % Among the 29 previously unknown resource leaks in open-source projects, 7 are confirmed by the project developers and PRs are accepted.
\end{itemize}

\section{Motivating Example}\label{sec:motivating}

This section presents a motivating example to underscore the significance of directly inferring resource-oriented intentions in code through a combination of background knowledge of resource management and code context understanding, rather than relying solely on mechanical matching of predefined resource acquisition/release APIs and \inlinecode{null}-checking conditions. The resource-oriented intentions refer to the functionalities of resource acquisition, release, and reachability validation involved in code.

\begin{figure*}
  \centering
  \includegraphics[width=1.85\columnwidth,scale=0.1]{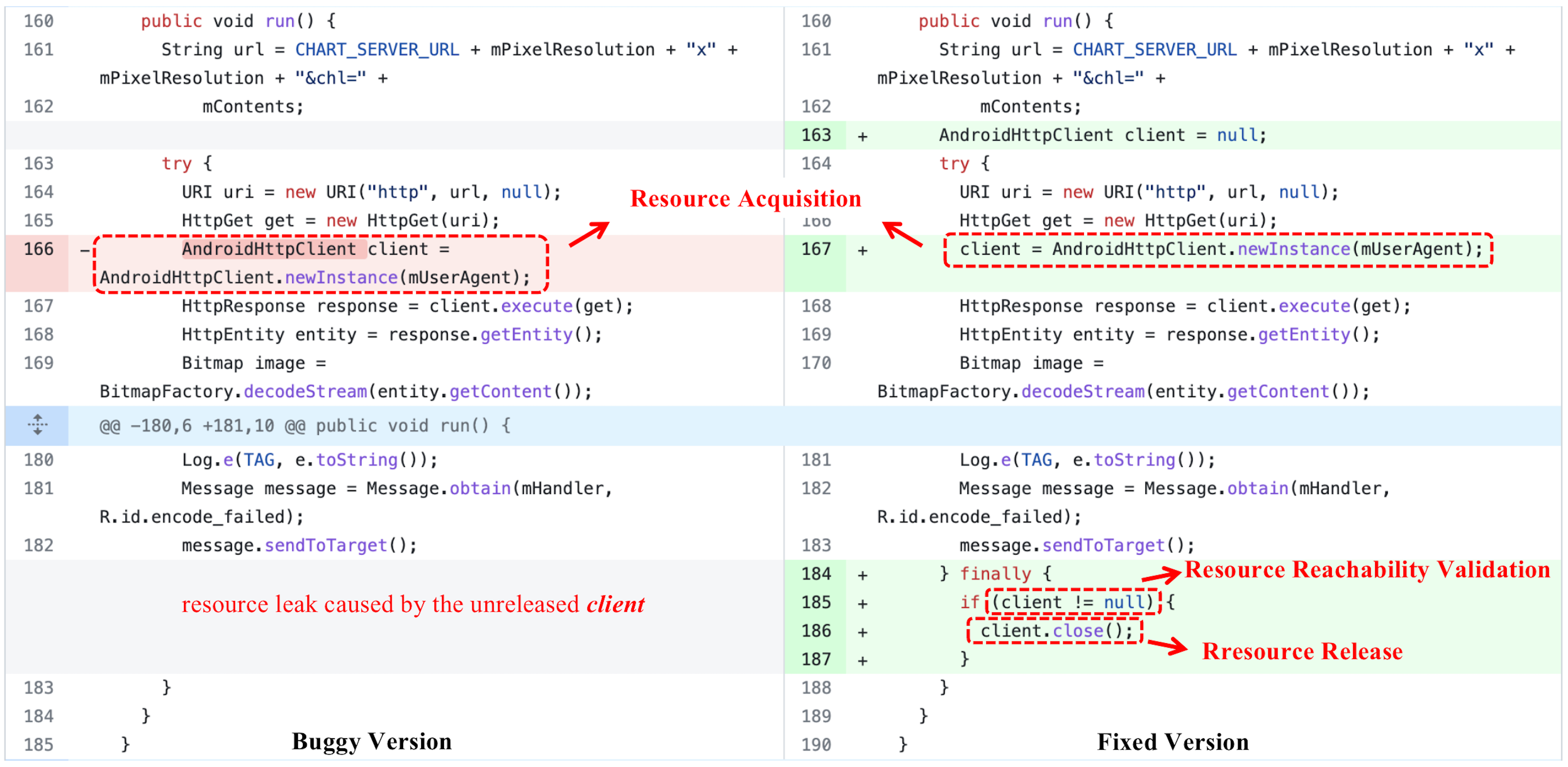}
  \vspace{-2mm}
  \caption{Comparison between Buggy Version (Left) and Fixed Version (Right) Associated with \inlinecode{AndroidHttpClient} Leak}
  \label{fig:motivating}
  \vspace{-5mm}
\end{figure*}

Figure~\ref{fig:motivating} illustrates the code diff of a commit~\cite{codediff} of fixing a resource leak associated with \inlinecode{AndroidHttpClient}, as observed in the DroidLeaks dataset. In the buggy version, the API \inlinecode{AndroidHttpClient.newInstance(...)} is called to acquire an \textit{HTTP client} resource at line 166, but the corresponding release API \inlinecode{AndroidHttpClient.close()} is not called, leading to a resource leak. In the fixed version, a \inlinecode{finally} structure is introduced at lines 184-187 to ensure the proper release of the acquired \textit{HTTP client} resource.
Through this example, we demonstrate the following key insights:

\textbf{1. Identifying resource acquisition/release more comprehensively.}
In the aforementioned example, the presence of a resource leak related to the \inlinecode{AndroidHttpClient} class goes unnoticed by all the eight subject detectors (i.e., Code Inspection~\cite{codeinspection}, Infer~\cite{infer}, Lint~\cite{lint}, SpotBugs~\cite{spotbugs}, Relda2-FS~\cite{wu2016relda2}, RElda2-FI~\cite{wu2016relda2}, Elite~\cite{liu2016understanding}, and Verifier~\cite{vekris2012towards}) evaluated in the DroidLeaks dataset. The reason for this oversight lies in the fact that \inlinecode{AndroidHttpClient} is not included in the predefined API sets used by these detectors. This limitation highlights a fundamental constraint of the existing detection paradigm, which primarily relies on locating API calls for resource acquisition/release through the resolution of API signatures in code (i.e., type inference) and subsequent matching with predefined API names.

However, by leveraging background knowledge encompassing pertinent concepts such as \textit{networking} and \textit{HTTP client}, and diligently analyzing the code contexts, which involve relevant terms like \inlinecode{URL} and \inlinecode{HTTP response}, it becomes evident that the API call \inlinecode{AndroidHttpClient.newInstance(...)} is specifically intended to acquire an \textit{HTTP client} resource. This realization motivates the exploration of inferring resource acquisition and release intentions based on code semantics (e.g., the API calls and the surrounding contexts), enabling more comprehensive identification of resource leaks compared to conventional mechanical API matching.

Furthermore, in addition to \inlinecode{AndroidHttpClient}, a multitude of APIs are available in various libraries/packages that implement the functionality of \textit{HTTP client}. For instance, within the \textit{selenium} library~\cite{selenium}, there exist related APIs like \inlinecode{HttpClient}, \inlinecode{JdkHttpClient}, \inlinecode{NettyClient}, \inlinecode{TracedHttpClient}, \inlinecode{NettyDomainSocketClient}, \inlinecode{PassthroughHttpClient}, among others. The eight detectors evaluated in DroidLeaks also fail to detect resource leaks across all of them. This failure can be attributed to the detectors' limited capability of capturing the underlying common characteristics shared by these APIs, which essentially represent the same resource concept, namely, an \textit{HTTP client}, and its usage. The presence of this limitation further emphasizes the necessity of incorporating background knowledge and code understanding in resource acquisition and release identification.

\textbf{2. Identifying resource reachability validation more comprehensively.}
The validation of resource reachability is a crucial factor in determining the occurrence of resource leaks. 
\revise{Note that the meanings between \textit{reachable}/\textit{reachability} and the related terms (e.g., \textit{open}/\textit{closed}/\textit{valid}) are different. \textit{Open/closed/valid} describe the status of a resource itself; while \textit{reachable}/\textit{reachability} indicate whether the open status of a resource can be propagated to \textit{a certain path, branch, or statement}.  For example, in the fixed version in  Figure~\ref{fig:motivating}, the \inlinecode{if}-condition at line 185 validates the \textit{reachability} of the resource ``HTTP client'' by checking whether the object of client is null. As a result, the client resource is actually unreachable on the false branch of this \inlinecode{if}-condition, that is, the open status of the client resource cannot reach the false branch. If a resource leak detection technique ignores the \textit{reachability} of the client resource, it would cause a false alarm in the false branch of this \inlinecode{if}-condition. Therefore, reachability is essential for reducing false alarms in static resource leak detection.}

Existing detection techniques, such as Torlak et al.~\cite{torlak2010effective} and Cherem et al.~\cite{cherem2007practical}, typically identify reachability validation by matching \inlinecode{null}-checking conditions. While this approach provides a simple and intuitive means to recognize the reachability validation at line 185, it may overlook certain validation intentions present in other code. For specific resource types, dedicated APIs, such as \inlinecode{isClosed()} and \inlinecode{isDisabled()}, are employed to perform reachability validation. In a similar manner to resource acquisition/release identification, a more comprehensive identification of such validation intentions necessitates the incorporation of background knowledge pertaining to common validation verbs, in conjunction with a deep understanding of the code contexts, such as the \inlinecode{if}-conditions. By considering these elements, resource reachability validation can be assessed more comprehensively, enabling a more accurate detection of resource leaks in code.

\section{Approach}
Based on these insights, we propose \textbf{\textit{\app}}, a novel approach that directly \textbf{Infers} \textbf{R}esource-\textbf{O}riented \textbf{I}ntentions in code to boost static resource leak detection, by leveraging the power of large language models (LLMs). In the approach, LLMs function as extensive knowledge bases with exceptional code understanding capabilities~\cite{openai2023gpt4,bubeck2023sparks,petroni2019language,huang2022prompt}, rendering them well-suited for inferring resource-oriented intentions.

\begin{figure}
    \centering
    \includegraphics[width=1.0\columnwidth]{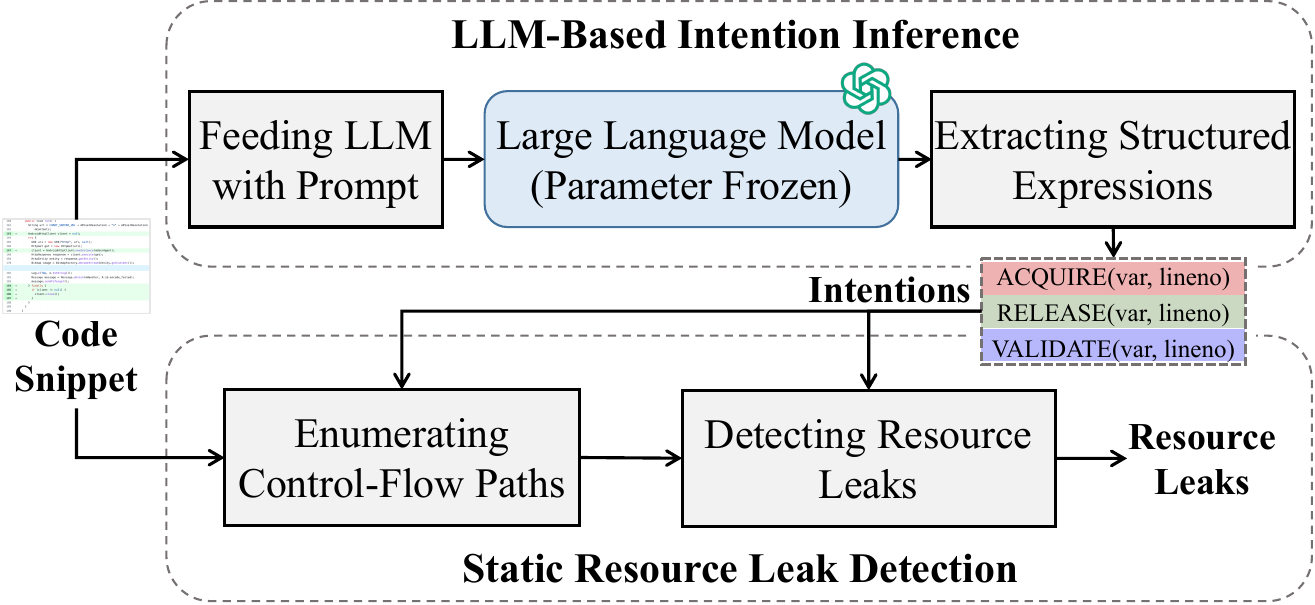}
    \vspace{-7mm}
    \caption{Approach Overview of \app}
    \label{fig:overview}
\end{figure}

Figure~\ref{fig:overview} represents the approach overview of \app.
For a given code snippet, \app first instantiates the prompt template to instruct the LLM in inferring the involved resource acquisition, release, and reachability validation intentions, which does no require any prior information on resource API pairs.  The intentions generated by LLMs (which are described in natural language) are then \revise{converted into structured expressions}, so as to facilitate the subsequent static analysis stage. By aggregating these inferred intentions, \app then proceeds to apply a two-stage static analysis algorithm to analyze the control-flow paths extracted from the code. The algorithm first identifies potential leak-risky paths based on the inferred resource acquisition/release APIs, and then prunes leak-risky paths with unreachable resources based on the inferred reachability validation across these paths on sibling branches, so as to reduce false alarms of resource leak detection.

\subsection{Resource-Oriented Intentions}
We focus on three resource-oriented intentions involved in code, which play a crucial role in resource management.

\textbf{Resource Acquisition.}
Statements acquire resources (e.g., locks or database connections), which are normally API invocations and involved with objects/variables of resources (e.g., \inlinecode{lock} or \inlinecode{db}). To structurally express the intentions of resource acquisition, we introduce the notation $\bm{ACQ(var, lineno)}$, where $var$ denotes the variable storing the acquired resource, and $lineno$ indicates the line number containing the statement responsible for acquiring the resource.
    
\textbf{Resource Release.}
Likewise, statements release previously acquired resources. We \revise{structurally represent} the intentions of resource release as $\bm{REL(var, lineno)}$, where $var$ corresponds to the variable storing the resource to be released, and $lineno$ indicates the line number of the statement responsible for performing the release operation.
    
\textbf{Resource Reachability Validation.}
Conditional statements aim to validate whether acquired resources remain reachable within specific branches. To \revise{structurally represent} the intentions of resource reachability validation, we employ the notation $\bm{VAL(var, lineno)}$, where $var$ represents the variable storing the resource being validated, and $lineno$ indicates the line number of the statement responsible for performing the validation operation. 

% \begin{table*}
%     \centering
%     \caption{Three Resource-oriented Intentions}
%     \begin{tabular}{c|c|c}
%         \hline
%         \textbf{Intention} & \textbf{Description} & \textbf{Form} \\ \hline
%         Resource Acquisition  & acquire a resource and store it into a variable & $acquire(var, lineno)$ \\ \hline
%         Resource Release & release a acquired resource & $release(var, lineno)$ \\ \hline
%         Resource Reachability Validation & validate the reachability of a acquired resource & $validate(var, lineno)$ \\ \hline
        
%     \end{tabular}
% \end{table*}

\textbf{\textit{Example 3.1.}}
In the method depicted in Figure~\ref{fig:motivating}, the intentions of the API calls at line 167, which aims to acquire a \textit{HTTP client} resource, can be represented as $ACQ(\inlinecode{client}, 167)$. The intention of the API call at line 186, responsible for releasing the acquired \textit{HTTP client} resource, can be denoted as $REL(\inlinecode{client}, 186)$. The intention of the \inlinecode{if} statement at line 185, intended to validate the reachability of the acquired \textit{HTTP client} resource, can be expressed as $VAL(\inlinecode{client}, 185)$.
\subsection{LLM-Based Intention Inference}\label{sec:intent}
To infer resource-oriented intentions present in code, we first prompt the large language model (LLM) with designed templates and then convert the LLM-generated answers into \revise{structured} expressions for the subsequent analysis. 

\subsubsection{Feeding LLM with Prompt}\label{sec:prompt}
\begin{figure}
    \centering
    \includegraphics[width=1.0\columnwidth]{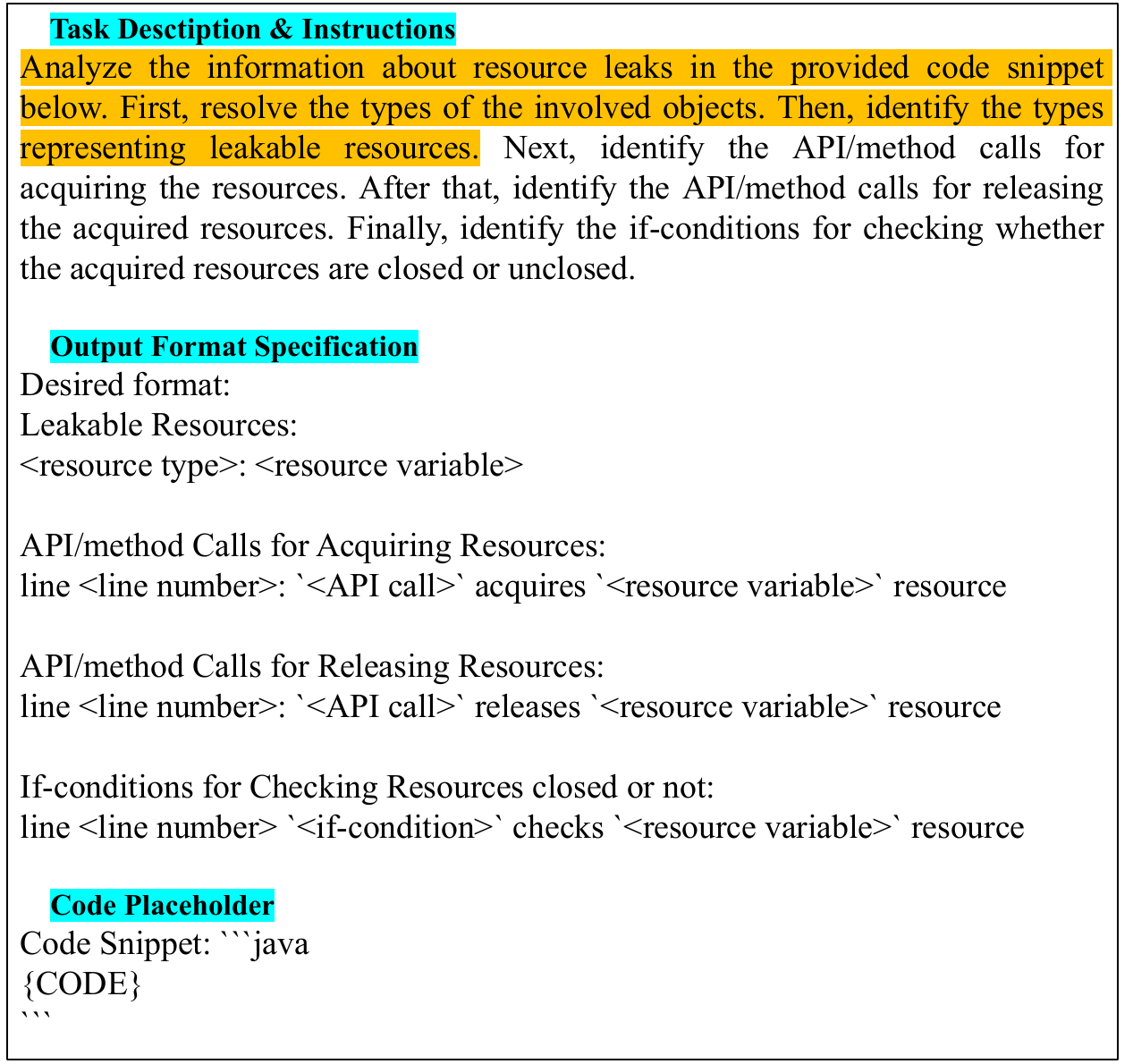}
    \vspace{-8mm}
    \caption{Prompt Template}
    \label{fig:prompt}
    \vspace{-6mm}
\end{figure}

For a given code, we construct a prompt using the template depicted in Figure~\ref{fig:prompt}. This prompt is then fed into the LLM (e.g., \gptfour) for intention inference. The prompt template is designed following the best practices for prompt engineering by OpenAI~\cite{openai_best}, consisting of three main components: Task Description \& Instructions, Output Format Specification, and Code Placeholder.

The \textbf{Task Description \& Instructions} \revise{component consists of several sequential instructions designed based on the chain-of-thought (CoT) idea~\cite{wei2022chain}. It begins with three \colorbox[HTML]{FFC100}{leading instructions} to introduce the overall task goal and guide the LLM in forming a smooth reasoning chain from a basic starting point, specifically by performing type inference for the involved objects and determining which types represent leakable resources.} 
Based on the analysis, the LLM is then required to infer the three types of resource-oriented intentions by the follow-up three instructions. Throughout the analysis and inference process, the LLM leverages its extensive background knowledge to understand the code contexts and execute the instructions effectively.

\textbf{Output Format Specification} aims to specify the format of identified intentions, facilitating subsequent parsing.

\textbf{Code Placeholder} is where the given code with line numbers is inserted, creating a specific prompt.

\begin{figure}
    \centering
    \includegraphics[width=1.0\columnwidth]{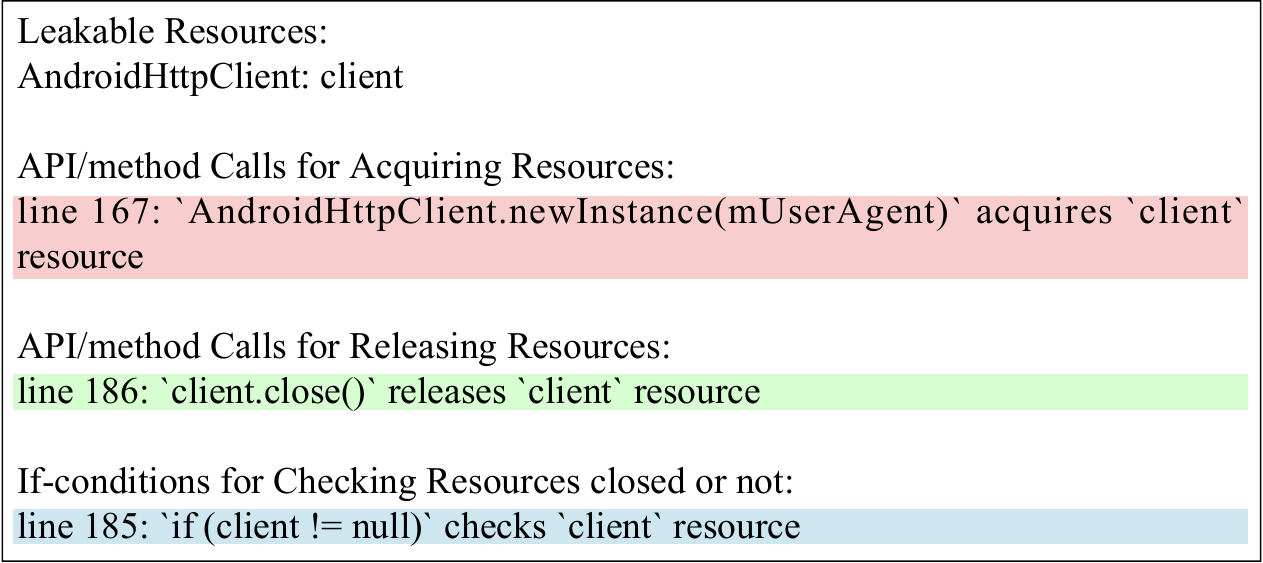}
    \vspace{-8mm}
    \caption{Answer Generated by \gptfour Model}
    \label{fig:answer}
\end{figure}

\textbf{\textit{Example 3.2.}}
Applying the template, we construct a prompt for the fixed version (right) illustrated in Figure~\ref{fig:motivating}. We then employ the \gptfour model~\cite{gpt-4} to generate an answer, as shown in Figure~\ref{fig:answer}. Notably, the \gptfour model exhibits the capability to effectively analyze code semantics and discern resource-oriented intentions. Within the generated answer, resource acquisition intentions are highlighted in \acqbox{red}, resource release intentions in \relbox{green}, and resource reachability validation intentions in \validbox{cyan}.

\subsubsection{Extracting \revise{Structured} Expressions}

To extract \revise{structured} expressions of the intentions identified by the LLM, we employ matching patterns that correspond to the output format specification specified in the template. Given an LLM-generated answer, we utilize regular expressions such as ``line <line number>: <API call> acquires <resource variable> resource'' to \revise{automatically} extract the line number, API call, and resource variable involved in resource acquisition intentions. Similar regular expressions can be used to extract information related to resource release intentions and reachability validation intentions. We then use the extracted information to instantiate the \revise{structured} expressions of the three types of intentions.

\textbf{\textit{Example 3.3.}}
From the answer illustrated in Figure~\ref{fig:answer}, the three highlighted lines are matched by the matching patterns. Subsequently, the three \revise{structured} intention expressions are extracted and instantiated as $ACQ(\inlinecode{client}, 167)$, $REL(\inlinecode{client}, 186)$, and $VAL(\inlinecode{client}, 185)$.

\subsection{Static Resource Leak Detection}
We then propose a lightweight static analysis based on the inferred resource-oriented intentions to detect resource leaks.

\subsubsection{Enumerating Control-Flow Paths}
Similar to most existing resource leak detection methods~\cite{torlak2010effective,wu2016light,spotbugs,infer}, our detection approach is also built upon control-flow analysis. We construct control-flow graphs (CFGs) for the code to detect and extract pruned control-flow paths from these graphs.
%Given a method-level code snippet, we begin by parsing it into an abstract syntax tree (AST) and then proceed to construct a CFG by traversing the AST nodes and inserting various control-flow edges between the code statements. The procedure is implemented based on the top of an established CFG construction tool~\cite{progex}. \yiling{I move this to the implmentation as suggested by reviewers}
\revise{Within the CFG, nodes that generate multiple subsequent control-flow branches, such as \inlinecode{if}-statements and \inlinecode{switch}-statements, are referred to as \inlinecode{branch}-nodes.} Next, we enumerate the paths within the CFG, considering the paths between the \inlinecode{entry}-node and \inlinecode{exit}-node. To ensure the efficiency of the enumeration process, we employ two pruning operations.

\textbf{Loop Structures.} Loop statements in the CFG can introduce circular paths, leading to infinite-length paths during the enumeration. To prevent this, we adopt a strategy that involves expanding the true-branch of a loop only once. 
% For instance, for a \inlinecode{while}-loop, we visit the statement in its true-branch once and then proceed directly to the end of the loop.

\textbf{Resource-Independent Branches.} The presence of branches in the CFG can exponentially increase the number of paths during enumeration, but most of the branches are not related to resource management. 
To address this, for a \inlinecode{branch}-node, we further prune its branches that are independent of resource acquisition/release and procedural exit (e.g, \inlinecode{return}-statements).
% To address this, for a \inlinecode{branch}-node, we further examine whether all of its branches are independent of resource acquisition/release and procedural exit (e.g, \inlinecode{return}-statements). 
% If the conditions are met, we only retain one branch for the \inlinecode{branch}-node, as all the other branches exhibit the same resource acquisition/release behaviors. The examination involves the inferred resource-oriented intentions, which help determine whether a CFG node acquires/releases resources.

\textbf{\textit{Example 3.4.}}
For the buggy version in Figure~\ref{fig:motivating}, there is only one control-flow path that sequentially executes from line 160 to line 185. In the fixed version shown in Figure~\ref{fig:motivating}, the \inlinecode{if}-statement for reachability validation at line 185 introduces two control-flow paths, corresponding to the line ranges [160-185, 186, 187-190] and [160-185, 187-190].

\subsubsection{Detecting Resource Leaks}
After enumerating the control-flow paths and inferring the resource-oriented intentions, \app proceeds with resource leak detection.

For each concerned resource involved in the inferred intentions, the detection process involves a two-stage path analysis, as illustrated in Algorithm~\ref{alg:detection}. This algorithm takes the concerned resource $res$, all the control-flow paths $\mathcal{P}$, and the resource-oriented intention set $\mathcal{I}$ as inputs.
The first stage (lines 1-16) is a \textit{single-path analysis} that initially identifies leak-risky paths of the concerned resource based on the $ACQ$ and $REL$ intentions.
The second stage (lines 17-29) is a \textit{cross-path analysis} that greedily eliminates the false-alarm-introducing \inlinecode{branch}-nodes by considering the $VAL$ intentions.
After completing both stages, the algorithm (lines 31-36) checks if any path has a corresponding status of $true$. If such a path exists, a resource leak of $res$ is reported.

\begin{algorithm}[t]
    \caption{Leak Detection via Two-Stage Path Analysis}
    \label{alg:detection}
    \small
    \begin{algorithmic}[1]
        \Require 
            $res$: concerned resource;
            $\mathcal{P}$: paths;
            $\mathcal{I}$: intention set;
        % \State $status\_map \gets \{\}$; 
        % \State $prefix\_map \gets \{\}$;  

        \Statex \LeftComment{\textcolor{blue}{Stage 1: \textit{Single-Path Analysis}}}
        % \State $status\_map \gets \{\}$; 
        \For{$path \in \mathcal{P}$} 
            \State $rd\_counter \gets 0$;  %\Comment{counter of resource descriptor}
            \For{$node \in path$}
                \State $ln \gets$ get line number of $node$;
                \If{$ACQ(res, ln) \in \mathcal{I}$}
                    \State $rd\_counter \gets rd\_counter + 1$;
                \ElsIf{$REL(res, ln) \in \mathcal{I}$}
                    \State $rd\_counter \gets rd\_counter - 1$;
                % \ElsIf{$VAL(res, ln) \in \mathcal{I}$}
                %     \State $prefix \gets$ get subpath of $path$ before $node$;
                %     \If{$prefix$ not in $prefix\_map$}
                %         \State $prefix\_map[prefix] \gets [path]$; 
                %     \Else
                %         \State $prefix\_map[prefix].push(path)$
                %     \EndIf
                \EndIf
            \EndFor
            \If{$rd\_counter > 0$}  %\Comment{potentially leaked or not}
                \State $path.risky \gets$ \textbf{true};   
            \Else
                \State $path.risky \gets$ \textbf{false};
            \EndIf
        \EndFor

        \Statex \LeftComment{\textcolor{blue}{Stage 2: \textit{Cross-Path Analysis}}}
        \State $br\_nodes \gets$ get all \inlinecode{\footnotesize branch}-nodes in the paths;
        \State sort $br\_nodes$ by their line numbers in a descending order
        \For{$br\_node \in br\_nodes$}  % \Comment{\textcolor{blue}{\textit{Status Back-Propagation}}}
            \State $br\_ln \gets$ get line number of $br\_node$
            \If{\textbf{not} $VAL(res, br\_ln) \in \mathcal{I}$}
                \State \textbf{continue};
            \EndIf
            \State $G_{prefix} \gets$ group paths containing $br\_node$ by path prefixes;
            \For{$g \in G_{prefix}$}  
                \State $(B_1, B_2, ...) \gets$ group paths in $g$ by branches of $br\_node$;
                \State $\Call{Propagate}{B_1, B_2, ...}$; 
                % \If{$status\_map[path] | \forall path \in B_1$ \textbf{and} all $path \in B_1$}
                %     \State $PROPAGATE(B_1, B_2, status\_map)$;
                % \ElsIf{$status\_map[path] = true$}
                %     \State 
                % \EndIf
            \EndFor
        \EndFor
        % \LeftComment{step (ii): \textit{cross-path analysis}
        \Statex \LeftComment{\textcolor{blue}{\textit{Reporting Leaks}}}
        \For{$path \in \mathcal{P}$}
            \If{$path.risky$}
                \State report a leak of $res$; 
                \State \textbf{break};   
            \EndIf
        \EndFor
    \end{algorithmic}
    % \vspace{3mm}
    \begin{algorithmic}[1]
        \Procedure{Propagate}{$B_1, B_2, ...$}
            \If{($\forall p \in B_i \mapsto p.risky$) \textbf{and} ($\nexists p \in \underset{j\neq i}{\cup} B_j \mapsto p.risky$)}
                \State foreach $p \in B_i: p.risky \gets$ \textbf{false};
            % \ElsIf{($\nexists p \in B_1 \mapsto p.risky$) \textbf{and} ($\forall p \in B_2 \mapsto p.risky$)}
            %     \State foreach $p \in B_2: p.risky \gets$ \textbf{false};
            \EndIf
        \EndProcedure
    \end{algorithmic}
\end{algorithm}

\textbf{Stage 1: Single-Path Analysis.}
The algorithm conducts an iterative process, sequentially processing each $path$ in $\mathcal{P}$ and traversing the nodes within $path$ (lines 2-17). During the traversal, a descriptor counter $rd\_counter$ keeps track of the $ACQ$ and $REL$ intentions involving $res$ within $path$ (lines 3-11). If $rd\_counter$ is greater than 0, it indicates that $path$ is a potential leak-risky path for $res$, resulting in the risky status of $path$ (i.e., $path.risky$) is initialized as $true$ (lines 12-16). \revise{When no released API is identified, there is no $REL$ item in the inferred intention expression. For example, when there is an $ACQ$ in a path but no corresponding $REL$, the rd\_counter will be 1, marking the path as risky (Line 12).}

\textbf{Stage 2: Cross-Path Analysis.}
As outlined in Section~\ref{sec:motivating}, the presence of branches, denoted as unreachable branches, where the resource is inaccessible, gives rise to the possibility of false-positive (FP) leak-risky paths. This, in turn, results in false alarms being reported at line 32. During this phase, we conduct cross-path analysis utilizing the inferred reachability validation intentions to eliminate the false alarms.

Intuitively, addressing this problem involves specifically analyzing each \inlinecode{branch}-node intended for reachability validation and identifying its unreachable branch. However, the task is complicated by the diverse implementations of validation APIs, as discussed in Section~\ref{sec:motivating}. Some APIs (e.g., \inlinecode{isClosed()}) return true to indicate that the resources are reachable, while others (e.g., \inlinecode{isActive()}) indicate the opposite. To overcome this challenge, our approach utilizes a novel heuristic that collectively analyzes all branches of a \inlinecode{branch}-node  for reachability validation to assess whether it introduces false alarms. If affirmative, such \inlinecode{branch}-nodes are designated as \textit{false-alarm-introducing} (FAI) nodes. These FAI nodes are characterized by its one branch being \textit{completely} leak-risky, while the others are \textit{completely} not leak-risky. A branch is considered \textit{completely} leak-risky when all the paths belonging to it exhibit a risky status of $true$. The completely leak-risky branch is the unreachable branch, and the paths within it are FP leak-risky paths for $res$. The insight behind this decision is that if the other branch (the completely not leak-risky branch) is unreachable, it suggests the presence of more severe functional bugs rather than resource leaks, such as \textit{null pointer errors}. 

\revise{To illustrate, consider the if-condition node (line 185) in the Fixed Version in Figure~\ref{fig:motivating}. It validates the \inlinecode{client} resource and then releases it. When one branch is completely leak-risky (marked in Stage 1 due to not containing \inlinecode{client.close()}) and the other is completely not leak-risky (marked in Stage 1 due to containing \inlinecode{client.close()}), we can essentially determine that this if-condition is a FAI node and its completely leak-risky branch is the unreachable branch. Otherwise, if \inlinecode{client.close()} is called in an unreachable branch (e.g., \inlinecode{if (client == null) \{client.close()};\}), it would result in a more severe and obvious null-pointer exception.}

% It is essential to emphasize that the heuristic aims at removing false alarms by identifying the FAI nodes instead of identifying all FP leak-risky paths. When the unreachable branch of a reachability validation statement is \textit{partially} leak-risky, the heuristic will overlook some FP leak-risky paths in the unreachable branch. However, the overall correctness of determining whether the statement is a FAI node remains intact, as true-positive (TP) leaked paths are still present in the other branches.

% % the absolute soundness of 
% identifying all FP leak-risky paths, its primary objective is to effectively identify and remove FAI paths, thereby minimizing the risk of false alarms for the concerned resource ($res$). 

% The unsound scenario may arise when both branches are \textit{partially} leak-risky, and the \textproc{Update} procedure inadvertently overlooks some FP paths in the unreachable branch. However, the overall correctness of the final detection results remains intact, as true-positive (TP) leaked paths are still present in the other branch, ensuring that the reported resource leak of $res$ is not a false alarm. This approach strikes a balance between soundness and effectiveness in the context of static resource leak detection.

To implement this heuristic, our algorithm initiates by identifying all \inlinecode{branch}-nodes in the CFG and iterates through them from back to front (lines 18-30). During this process, each \inlinecode{branch}-node $br\_node$ is examined to determine if it involves a $VAL$ intention. If not, the algorithm skips $br\_node$, recognizing that it does not influence the risky status of the paths (lines 21-24). 
On the contrary, when $br\_node$ involves a $VAL$ intention, the algorithm conducts two grouping operations (line 24 and line 26, respectively). This results in the division of paths containing $br\_node$ into branches $(B_1, B_2, ...)$. All these branches share a common path prefix (a subpath from the \inlinecode{entry}-node to $br\_node$) but belong to different branches of $br\_node$. This implies that, at a specific execution path to $br\_node$, $(B_1, B_2, ...)$ represent potential subsequent branches of $br\_node$. Upon the branches, the algorithm performs the heuristic by executing the \textproc{Propagate} procedure and updating the risky status of FP leak-risky paths to $false$ (line 28).

% It is essential to emphasize that while this stage does not guarantee 
% % the absolute soundness of 
% identifying all FP leak-risky paths, its primary objective is to effectively identify and remove FAI paths, thereby minimizing the risk of false alarms for the concerned resource ($res$). 

% The unsound scenario may arise when both branches are \textit{partially} leak-risky, and the \textproc{Update} procedure inadvertently overlooks some FP paths in the unreachable branch. However, the overall correctness of the final detection results remains intact, as true-positive (TP) leaked paths are still present in the other branch, ensuring that the reported resource leak of $res$ is not a false alarm. This approach strikes a balance between soundness and effectiveness in the context of static resource leak detection.

\textbf{\textit{Example 3.5.}}
In the buggy version depicted in Figure~\ref{fig:motivating}, the detection algorithm successfully identifies a resource leak of the variable \inlinecode{client}. 
% It detects a path that leads to the leak, thus raising an alert for the presence of a resource leak involving \inlinecode{client}.
In contrast, for the fixed version shown in Figure~\ref{fig:motivating}, the algorithm identifies a leak-risky path [160-185, 187-190] of the variable \inlinecode{client} in Stage 1. However, the \textproc{Update} procedure is then executed in Stage 2, which effectively eliminates this path as an FAI path. Consequently, no resource leak of \inlinecode{client} is reported in this version, indicating that the detection algorithm correctly avoids false alarms and provides a more accurate analysis of resource leaks.

\section{Implementation}
The current implementation of \app utilizes the widely adopted \gptfour as the LLM and sets the model temperature as 0 to remove generation randomness.
\revise{For RQ4, we use \textit{Meta-Llama-3-8B-Instruct}~\cite{Llama3-HF} and \textit{gemma-2-9b-it}~\cite{Gemma2-HF} versions available on the Hugging Face Platform, and run the model inference with the Transformers library~\cite{Transformers}. To align with the temperature setting (i.e., temperature=0) of \gptfour, we use greedy search decoding for both models, ensuring there is no randomness. The experiments are run on a workstation with 4 AMD EPYC 9554 64-Core CPUs, 512 GB of memory, and a single Nvidia H100 (80GB) GPU.}
\revise{For the CFG construction in static resource leak detection, given a method-level code snippet, we begin by parsing it into an abstract syntax tree (AST) and then proceed to construct a CFG by traversing the AST nodes and inserting various control-flow edges between the code statements. The procedure is implemented based on the top of an established CFG construction tool~\cite{progex}.}
\cameraready{Since some resources are intentionally kept open and returned for further use, we adopt a rule to exclude the leak reports where the inferred resource objects returned in \inlinecode{return}-statements, thereby reducing false positives.}
In addition, we take into account certain Java-specific language features, such as \inlinecode{try-with-resources} statements.
% such as \inlinecode{try-(catch)-finally} structures and \inlinecode{try-with-resources} statements.
% Regarding the \inlinecode{try-(catch)-finally} structure, the statements within the \inlinecode{finally} block are executed in all control-flow paths. As a result, when enumerating the control-flow paths on control-flow graphs (CFGs), we further pay special attention to processing the statements within the \inlinecode{finally} block to ensure the paths are complete.
For the \inlinecode{try-with-resources} statements, the acquired resources are automatically closed, and there is no need for manual resource closure~\cite{try-with-resources}. To address this, we incorporate a rule-based post-processing step to filter out false alarms caused by \inlinecode{try-with-resources} statements. \revise{Specifically, for each reported resource leak, we traverse the AST of the examined method to determine if the leaked resource is acquired within a \inlinecode{try-with-resources} statement. If so, the reported leak is omitted.}
\section{Evaluation}

We conduct comprehensive experiments to evaluate the effectiveness of \textit{\app} in detecting resource leaks by employing the DroidLeaks dataset~\cite{liu2019droidleaks} and the JLeaks dataset~\cite{icse/Jleaks}.

\revise{
First, we apply \app to detect resource leaks in the code snippets from DroidLeaks and JLeaks datasets, comparing its effectiveness against baseline detectors (\textbf{RQ1}).
Next, we evaluate the effectiveness of \app in detecting previously unknown resource leaks in real-world open-source projects (\textbf{RQ2}).
We then conduct an ablation study to evaluate the contributions of intention inference and static analysis in \app (\textbf{RQ3}).
Subsequently, we investigate the generalizability of integrating \app with different LLMs and compare their effectiveness (\textbf{RQ4}).
Finally, we manually annotate the resource-oriented intentions in code to measure how effectively \app infers these intentions (\textbf{RQ5}).
}
% Finally, we manually annotate the resource-oriented intentions presented in code to measure the extent to which \app successfully infers these intentions (\textbf{RQ3.a}), and conduct an ablation study to analyze and compare the contributions of two key components of \app (\textbf{RQ3.b}).

All the research questions are listed as follows.

% \begin{itemize}[leftmargin=15pt]
% \item \textbf{RQ1 (Effectiveness in Leak Detection)}: How effectively can \app detect resource leaks in the DroidLeaks and JLeaks datasets? Can it outperform baseline detectors in this regard?
% \item \textbf{RQ2 (Effectiveness in Project Scanning)}: How effective is \app in detecting previously-unknown resource leak bugs when applied to open-source projects?
% \item \textbf{RQ3.a (Accuracy in Intention Inference)}: To what degree can \app accurately infer the resource-oriented intentions in code? 
% \item \textbf{RQ3.b (Ablation Study)}: What is the contribution of the intention inference and static analysis in \app?
% \end{itemize}

\begin{itemize}[leftmargin=*]
\item \textbf{RQ1 (Effectiveness in Leak Detection)}: How effectively can \app detect resource leaks in existing  benchmarks? 
\item \textbf{RQ2 (Effectiveness in Project Scanning)}: How effective is \app in detecting previously unknown resource leak bugs when applied to open-source projects?
\item \textbf{RQ3 (Ablation Study)}: What are the contributions of intention inference and static analysis in \app?
\item \revise{\textbf{RQ4 (Generalizability with Different LLMs)}: How generalizable is \app when integrated with different LLMs?}
\item \textbf{RQ5 (Accuracy in Intention Inference)}: How accurately can \app infer resource-oriented intentions in code?
\end{itemize}

\subsection{Effectiveness in Resource Leak Detection (RQ1)}
In this evaluation, we assess the effectiveness of \app with \gptfour in detecting resource leaks using two datasets.

\subsubsection{Data}
We utilize two datasets, namely \textbf{DroidLeaks}~\cite{liu2019droidleaks} and \textbf{JLeaks}~\cite{icse/Jleaks}, constructed by analyzing commits dedicated to fixing resource leak bugs. Each bug in the datasets is associated with a concerned resource type, which is determined by reviewing the corresponding bug-fixing commit. By retrieving the bug-fixing commit and the previous commit, both the buggy and fixed versions are obtained for each bug.

\begin{itemize}[leftmargin=15pt]
    \item \textbf{DroidLeaks}: This dataset comprises resource leak bugs sourced from open-source Android applications. Originally, the DroidLeaks dataset included 116 bugs used to assess the performance of eight widely-used resource leak detectors~\cite{codeinspection,infer,lint,spotbugs,wu2016relda2,liu2016understanding,vekris2012towards}. In our evaluation, we attempt to acquire all 116 bugs, successfully obtaining both the buggy and fixed versions for 86 of them. Consequently, the final collection comprises 172 code snippets covering 28 different resource types, such as \inlinecode{Cursor}.

    \item \textbf{JLeaks}: This is a more recent dataset containing resource leak bugs from open-source Java projects. The original JLeaks dataset included 1,094 bugs used to evaluate the performance of three widely-used resource leak detectors~\cite{spotbugs,infer,pmd}. For our evaluation, we filtered out instances based on two conditions: 1) the resource type is annotated, and 2) both the buggy and fixed versions are free of syntax errors. This filtering process resulted in 784 bugs (1,568 code snippets) encompassing 368 different resource types.
\end{itemize}

\subsubsection{Baselines}
\revise{In RQ1, we compare \app with three well-maintained industrial static detectors, as they have been evaluated in both DroidLeaks and JLeaks papers~\cite{liu2019droidleaks,icse/Jleaks}.}
\begin{itemize}[leftmargin=15pt]
    \item \textbf{SpotBugs}~\cite{spotbugs}: SpotBugs, formerly known as FindBugs, is a static analysis tool that identifies bugs and potential issues in Java programs.
    \item \textbf{Infer}~\cite{infer}: Infer is a static analysis tool developed by Facebook that detects various types of programming errors in C, C++, and Java code.
    \item \textbf{PMD}~\cite{pmd}: PMD is a widely-used static code analysis tool that identifies potential issues in code.
\end{itemize}
% , namely Code Inspection~\cite{codeinspection}, Infer~\cite{infer}, Lint~\cite{lint}, SpotBugs~\cite{spotbugs}, Relda2-FS~\cite{wu2016relda2}, RElda2-FI~\cite{wu2016relda2}, Elite~\cite{liu2016understanding}, and Verifier~\cite{vekris2012towards}

\subsubsection{Procedure}
For each bug in our experiments, we apply \app to both its buggy version and its fixed version. 
\revise{
If \app reports a resource leak in the buggy version related to the concerned resource type, it is considered a \textit{true positive} (TP); otherwise, there is a \textit{false negative} (FN). Conversely, if \app reports a resource leak in the fixed version for the corresponding resource type, it is a \textit{false positive} (FP). Based on this, we count the numbers of TPs, FPs, and FNs, and calculate \textit{Bug Detection Rate} (BDR) and \textit{False Alarm Rate} (FAR) following a previous study~\cite{liu2019droidleaks}.
$$ BDR = \frac{\text{\# TP}}{\text{\# Buggy Versions}}, \quad FAR = \frac{\text{\#FP}}{\text{\# Fixed Versions}} $$
Additionally, to better reflect the overall effectiveness of \app, we calculate the $F1$ score based on \textit{Precision} and \textit{Recall}:
$$ F1 = \frac{2 \cdot \textit{Precision} \cdot \textit{Recall}}{\textit{Precision} + \textit{Recall}}, $$
where \textit{Precision} and \textit{Recall} are calculated by $\sfrac{\text{\#TP}}{(\text{\#TP} + \text{\#FP})}$ and $\sfrac{\text{\#TP}}{(\text{\#TP} + \text{\#FN})}$, respectively.
Note that we do not report \textit{Precision} and \textit{Recall} separately in the experimental results, as their implications are effectively captured by \textit{Bug Detection Rate} and \textit{False Alarm Rate}.
}

% To compare the performance of \app with the eight detectors (i.e., baselines) evaluated in DroidLeaks, we calculate the bug detection rate and false alarm rate for each baseline directly using the data from the original DroidLeaks dataset. 
% For fair comparison, we use the same set of the 86 bugs as the experimented bugs for both \app and all the eight baseline detectors, instead of specifically selecting different experimented bugs for each detector as done in the original DroidLeaks evaluation.

\subsubsection{Results}

The evaluation results for DroidLeaks and JLeaks are presented in Table~\ref{tab:detection-droidleaks} and Table~\ref{tab:detection-jleaks}, respectively. 

% \begin{table}
% 	\centering
%     % \setlength{\tabcolsep}{0.2em}
%     % \setlength\extrarowheight{0.1em}
% 	\caption{Resource Leak Detection on DroidLeaks Dataset}\label{tab:detection-droidleaks}
%     \vspace{-2mm}
%     \renewcommand{\arraystretch}{1.2}
%     \begin{tabular}{@{}ccc@{}}
%         \toprule
%         \textbf{Detector} & \textbf{\makecell[c]{\# Detected Bugs \\(Bug Detection Rate)}} & \textbf{\makecell[c]{\# False Alarms \\(False Alarm Rate)}} \\ \midrule 
%         % Lint              & 10 (11.6\%)            & 0 (0.0\%)            \\
%         SpotBugs          & 6 (6.9\%)              & 0 (0\%)              \\
%         Infer             & 37 (43.0\%)            & 16 (18.6\%)          \\
%         PMD               & 10 (11.6\%)            & 6 (7.0\%)            \\
%         % Relda2-FS         & 11 (12.7\%)            & 9 (10.4\%)           \\
%         % RElda2-FI         & 8 (9.3\%)              & 4 (4.6\%)            \\
%         % Elite             & 6 (6.9\%)              & 4 (4.6\%)            \\
%         % Verifier          & 3 (3.4\%)              & 2 (2.3\%)            \\ \hline
%         % GPT               & 38 (44.1\%)            & 13 (15.1\%)          \\ 
%         \hline
%         \app (ours)       & 51 (59.3\%)            & 16 (18.6\%)            \\ \bottomrule
%     \end{tabular}
% 	% \end{adjustbox}
% \end{table}

\begin{table}
	\centering
    \setlength{\tabcolsep}{8pt}
	\caption{Resource Leak Detection on DroidLeaks Dataset}\label{tab:detection-droidleaks}
        \vspace{-2mm}
    % \vspace{-2mm}
    \renewcommand{\arraystretch}{1.2}
    \begin{tabular}{l|rr|rr|r}
        \Xhline{2\arrayrulewidth}
        \textbf{Detector} & \textbf{\#TP} & \textbf{\#FP} & \textbf{BDR}~$\uparrow$ & \textbf{FAR}~$\downarrow$ & \textbf{F1}~$\uparrow$ \\ 
        \hline 
        SpotBugs          & 6              & 0             & 6.9\%          & 0\%             & 0.129          \\
        Infer             & 37             & 16            & 43.0\%         & 18.6\%          & 0.532       \\
        PMD               & 10             & 6             & 11.6\%         & 7.0\%           & 0.196       \\
        \hline
        \app              & 51             & 16            & 59.3\%         & 18.6\%          & 0.667         \\ 
        \Xhline{2\arrayrulewidth}
    \end{tabular}
    % \begin{tabular}{l|cc|cc|cc|c}
    %     \toprule
    %     \textbf{Detector} & \textbf{\#TP} & \textbf{\#FP}   & \textbf{BDR}  & \textbf{FAR} & \textbf{Precision} & \textbf{Recall} & \textbf{F1} \\ 
    %     \midrule 
    %     SpotBugs          & 6             & 0               & 6.9\%         & 0\%          & 1.000    & 0.069     & 0.129          \\
    %     Infer             & 37            & 16              & 43.0\%        & 18.6\%       & 0.698    & 0.430     & 0.532       \\
    %     PMD               & 10            & 6               & 11.6\%        & 7.0\%        & 0.625    & 0.116     & 0.196       \\
    %     \midrule
    %     \app              & 51            & 16              & 59.3\%        & 18.6\%       & 0.761    & 0.593     & 0.667         \\ 
    %     \bottomrule
    % \end{tabular}
	% \end{adjustbox}
    \vspace{-3mm}
\end{table}

\begin{table}
	\centering
    \setlength{\tabcolsep}{8pt}
	\caption{\revise{Resource Leak Detection on JLeaks Dataset}}\label{tab:detection-jleaks}
        \vspace{-2mm}
    \renewcommand{\arraystretch}{1.2}
    \begin{threeparttable}
        \begin{tabular}{l|rr|rr|r}
            \Xhline{2\arrayrulewidth}
            \textbf{Detector} & \textbf{\#TP}  & \textbf{\#FP}  & \textbf{BDR}~$\uparrow$  & \textbf{FAR}~$\downarrow$ & \textbf{F1}~$\uparrow$ \\ 
            \hline 
            SpotBugs\tnote{*} & 44             & ---            & 24.3\%        & ---          & ---         \\
            Infer\tnote{*}    & 5              & ---            & 9.4\%         & ---          & ---         \\
            PMD               & 341            & 40             & 43.5\%        & 5.1\%        & 0.585       \\
            \hline
            \app              & 490            & 153            & 62.5\%        & 19.5\%       & 0.687        \\ 
            \Xhline{2\arrayrulewidth}
        \end{tabular}
        % \begin{tabular}{l|cc|cc|cc|c}
        %     \toprule
        %     \textbf{Detector} & \textbf{\#TP}  & \textbf{\#FP}  & \textbf{BDR}  & \textbf{FAR} & \textbf{Precision} & \textbf{Recall} & \textbf{F1} \\ 
        %     \midrule 
        %     SpotBugs\tnote{*} & 44             & -              & 24.3\%        & -            & -          & 0.243       & -           \\
        %     Infer\tnote{*}    & 5              & -              & 9.4\%         & -            & -          & 0.094       & -           \\
        %     PMD               & 341            & 40             & 43.5\%        & 5.1\%        & 0.895      & 0.435       & 0.585       \\
        %     \midrule
        %     \app              & 490            & 153            & 62.5\%        & 19.5\%       & 0.762      & 0.625       & 0.687        \\ 
        %     \bottomrule
        % \end{tabular}
    \begin{tablenotes}
            \item[*] \footnotesize{Due to encountering compilation issues, SpotBugs and Infer are able to process only 181 and 53 instances, respectively.}
        \end{tablenotes}
    \end{threeparttable}
	% \end{adjustbox}
\end{table}

\textbf{DroidLeaks:}
\app demonstrates a bug detection rate of 59.3\% with a false alarm rate of 18.6\% across 86 bugs in DroidLeaks. Compared to the baseline detectors, \app achieves notably higher bug detection, detecting 14\textasciitilde 45 more bugs, while maintaining a comparable false alarm rate. \revise{Considering the overall effectiveness, \app achieves an $F1$ score of 0.667, significantly outperforming the baselines by 25.4\% to 417.1\%.}

\textbf{JLeaks:}
\app shows a bug detection rate of 62.5\% with a false alarm rate of 19.5\% across 784 bugs in JLeaks. Compared to the baseline detectors, \app achieves significantly superior bug detection, identifying 149\textasciitilde 485 more bugs in JLeaks.
\revise{As JLeaks does not release complete compilable  project-level environments, we cannot run the baselines SpotBugs and INFER (which both require compiled projects) and Table~\ref{tab:detection-jleaks} presents their bug detection rate as reported in JLeaks paper. For the baseline PMD which does not require compiled projects, we run PMD on JLeaks and report both its false alarm rate and bug detection rate.} \revise{As shown in Table~\ref{tab:detection-jleaks}, although PMD achieves lower false alarm rate than \app, \app achieves much better trade-off between bug detection rate and false alarm rate than PMD, i.e., \app achieves an F1 score of 0.687, outperforming PMD by 17.4\%.}

%It's important to note that we do not report the false alarm rate of SpotBugs and Infer due to the unavailability of data (\revise{complete compilable/runnable project-level environments}) necessary for its calculation in current JLeaks dataset. \revise{Considering the overall effectiveness, \app achieves an $F1$ score of 0.687, significantly outperforming PMD with a 17.4\% improvement.}

\textbf{Analysis:} 
(i) \emph{Bug Detection Analysis}: The higher bug detection rate of \app can be attributed to the advantages of the resource-oriented intention inference utilized by \app, as discussed in the insights provided in Section~\ref{sec:motivating} (see RQ3.a for the resource coverage assessment). However, it is essential to acknowledge that the resource-oriented intention inference cannot entirely replace the value of more sound program analysis techniques. For instance, in the case of buggy versions not detected by \app, a substantial portion is caused by the intricacies of Android's lifecycle management and callback mechanisms. This observation indicates an important future direction, i.e., to explore better integration of LLM-based resource-intention inference with more advanced program analysis techniques, which could enable more general and powerful resource leak detection. 
(ii) \emph{False Alarm Analysis}: False alarms reported by \app primarily stem from incorrect or missing intentions. When \app incorrectly identifies an acquisition intention or overlooks a release intention, a false alarm occurs. To address this issue, we intend to explore the integration of LLM-based prompt engineering techniques, such as in-context learning, to enhance the capabilities of LLMs. Notably, a significant portion of these ``false alarms'' arise from incorrect ground truths. Since both datasets are curated based on bug-fixing commits, certain ``fixed'' versions still contain resource leaks. This underscores the ongoing challenge of ensuring the quality of resource leak datasets.

%which could cover a broader range of application scenarios and enhance the overall detection capabilities.

\finding{
    \app demonstrates a bug detection rate of 59.3\% with a false alarm rate of 18.6\% across 86 bugs in DroidLeaks, and a bug detection rate of 62.5\% with a false alarm rate of 19.5\% across 784 bugs in JLeaks. \revise{Considering overall effectiveness, \app significantly outperforms the baselines on both datasets, with $F1$ scores of 0.667 and 0.687.} This shows a promising potential of \app in detecting resource leaks.
    % rather than purely relying on mechanical matching of predefined resource acquisition/release APIs and \inlinecode{null}-checking conditions. 
    % These results underscore the remarkable effectiveness of \app in accurately detecting resource leaks, emphasizing its potential as a valuable tool for improving software quality and reducing the risk of resource-related defects.
}

\subsection{Effectiveness in Open-Source Project Scanning (RQ2)}
We apply \app with \gptfour on real-world Java projects to evaluate its efficacy in detecting previously-unknown resource leaks. 

\subsubsection{Data}
\revise{To save time and reduce the cost of querying the \gptfour API, we employ two strategies to sample two set of methods from open-source projects.}

\revise{\textbf{Suspicious Methods Filtered by Matching Resources.}}
We crawl 115 Java open-source projects with more than 50 stars, which are created after December 31, 2021. We randomly filter 100 suspicious methods from 13 projects, after matching 20 common resource terms reported in previous work~\cite{wang2023mining}, instead of completely scanning these projects. The employed terms are listed as follows: \textit{stream}, \textit{reader}, \textit{client}, \textit{writer}, \textit{lock}, \textit{player}, \textit{connection}, \textit{monitor}, \textit{gzip}, \textit{ftp}, \textit{semaphore}, \textit{mutex}, \textit{camera}, \textit{jar}, \textit{buffer}, \textit{latch}, \textit{socket}, \textit{database}, \textit{scanner}, \textit{cursor}. 

\revise{
\textbf{Random Methods Sampled from Apache Lucene.}
To further evaluate the effectiveness of \app in detecting resource leaks in the scenarios without relying on matching common resource terms, we also randomly sample 100 methods from the latest version of \textit{Apache Lucene}~\cite{Lucene}, a widely used text retrieval engine.
}

\cameraready{
\textbf{Interprocedural Scanning with CodeQL on Apache Lucene.}
To investigate the potential of leveraging inferred intentions to enhance interprocedural scanning, we integrate the \textit{Resource Acquisition} and \textit{Resource Release} intentions inferred by \app for Apache Lucene into CodeQL QL scripts. This integration is straightforward, as CodeQL supports the addition of custom resource types and their associated acquisition and release APIs. However, we have not yet incorporated \textit{Resource Reachability Validation}, as modifying CodeQL's flow analysis to support it presents significantly greater complexity.
}

\subsubsection{Procedure}
We apply \app on these 200 methods to detect resource leaks. For each reported leak, we first manually annotate whether it is a true bug by reading code and consulting related information. For the true bugs, we then submit the corresponding bug-fix pull requests and help the project developers review them.
\revise{We run PMD on these methods and manually check the reported bugs. In RQ2,  we use PMD instead of SpotBugs and Infer as our baseline,  because (i) PMD shows  the best detection rate on the large-scale benchmark JLeaks in RQ1, and (ii) PMD does not require compiled projects (while each instance in RQ2 is a method snippet).}

\begin{table}
	\centering
	\caption{Results of Open Source Scanning}\label{tab:open-source}
    \vspace{-2mm}
    \renewcommand{\arraystretch}{1.2}
    \begin{tabular}{l|lr|rr}
        \Xhline{2\arrayrulewidth}
        \multirow{2}{*}{\textbf{Detector}} & \multicolumn{2}{c|}{\textbf{100 Suspicious Methods}}       & \multicolumn{2}{c}{\textbf{100 Lucene Methods}} \\
        \cline{2-3}\cline{4-5}
                                           & \textbf{\#TP} & \textbf{\#FP}                          & \textbf{\#TP} & \textbf{\#FP} \\ 
        \hline 
        \app              & ~~26 (\textit{7 Accepted PRs}) & 3             & 3              & 0               \\ 
        PMD               & ~~18             & 2             & 0              & 0               \\
        \Xhline{2\arrayrulewidth}
    \end{tabular}
	% \end{adjustbox}
\end{table}

\subsubsection{Results}
In the 100 suspicious methods filtered by matching resources, \app reports 29 resource leaks and 26 are annotated as true bugs. At the submission time, 7 of the true bugs have been confirmed by the project developers, and the pull requests have been accepted.
PMD reports 20 resource leaks in the 100 methods, with 18 of them being true bugs. Among these 18 bugs identified by PMD, \app successfully detects 17 of them. The lone bug missed by \app involves a method comprising over 400 lines of code, posing challenges for LLM processing. Conversely, \app identifies 9 additional true bugs not caught by PMD. Noteworthy among these are resource types such as \inlinecode{ManagedBuffer} from the \textit{Apache Spark} library, \inlinecode{InMemoryDirectoryServer} from \textit{UnboundID LDAP}, and a custom type \inlinecode{JDBCConnection} specific to the project.

\revise{
In the 100 methods sampled from Apache Lucene, \app reports 3 resource leaks, all of which are confirmed as true bugs, while PMD reports zero bugs. The 3 bugs detected by \app involve project-defined resource classes \inlinecode{IndexReader}, \inlinecode{DirectoryReader}, and \inlinecode{DirectoryTaxonomyReader}. These resources are acquired and released by calling the \inlinecode{incRef} and \inlinecode{decRef} methods, respectively, which are challenging for static detectors to address.}

\cameraready{
For the interprocedural scanning, CodeQL reports 34 leaks before integrating \app-inferred intentions and 49 leaks after integration. Among these, 29 (85.3\%) and 41 (83.7\%) are confirmed as true bugs, respectively. 
}

% \textbf{Comparison with Resource Mining Approach:}
% Of the 26 bugs, 3 are beyond the scope of the state-of-the-art resource acquisition/release API pair (RAR pair) mining approach MiROK~\cite{sigsoft/WangL0LZ23}. MiROK primarily concentrates on extracting the \emph{definitions} of abstract RAR pairs, which differs from our objective of directly detecting resource-oriented intentions in the \emph{usages} of RAR pairs. Nonetheless, the abstract RAR pairs mined by MiROK can aid in filtering suspicious methods for \app, similar to the 20 resource terms employed in this evaluation.

\finding{
    \app successfully detects 26 previously unknown bugs in 100 suspicious methods and 3 previously unknown bugs in 100 Lucene methods. The results demonstrate the practical capability of \app in effectively detecting resource leaks in real-world software projects. \cameraready{Integrating \app-inferred intentions enables CodeQL to identify 12 additional true bugs during interprocedural scanning on Apache Lucene.}
}
\subsection{Ablation Study}
\revise{
We compare \app with three end-to-end GPT-based detectors that rely solely on prompting engineering to evaluate the contribution of static analysis. 
% Then, we apply a variant chain-of-thought (CoT) prompt in \app's intention inference to assess the impacts of different prompting strategies.
}

\subsubsection{Procedure}
The three end-to-end GPT-based detectors without static analysis are denoted as \gptleak variants (The used prompts can be found in our replication package~\cite{replication}).

\begin{itemize}[leftmargin=15pt]
    \item \textbf{\gptleak} employs a straightforward instruction to directly identify resource leaks in the provided code.
    \item \revise{\textbf{\gptleakexp} employs an additional chain-of-thought (CoT) instruction ``\textit{First, explain the behavior of the code}'' to explain code behavior, which have been proven effective in static bug detection in previous study~\cite{wen2024automatically}.}
    \item \revise{\textbf{\gptleakroi} employs additional CoT instructions to infer resource-oriented intentions (ROIs) in code, similar to \app, before identifying resource leaks.}
\end{itemize}

We apply the three detectors to the 172 code snippets collected from the DroidLeaks dataset and calculate the BDR, FAR, and $F1$ score for each detector.
% \revise{
% \noindent\textbf{\app Variants with Alternative CoT Prompts.}
% In \app, the leading instructions in the CoT prompt for intention inference (see Figure~\ref{fig:prompt}) can be modified. We create two variants of \app by changing these instructions. First, we replace them with empty content, resulting in the variant named \textbf{\appempty}. Second, we substitute the leading instructions with ``Initially, explain the behavior of the code'', as used in a previous study for static bug detection~\cite{wen2024automatically}, creating the variant \textbf{\appexp}.
% }

% contributions of two key components of \app: the LLM-based resource-oriented intention inference and the static resource leak detection.

% \begin{figure}
%     \centering
%     \includegraphics[width=1.0\columnwidth]{images/ablation-prompts.pdf}
%     \vspace{-7mm}
%     \caption{\revise{Prmopt Templates for the Three End-to-End GPT-based Detectors}}
%     \label{fig:ablation-prompts}
%     \vspace{-3mm}
% \end{figure}

\subsubsection{Results}
\revise{
As shown in Table ~\ref{tab:ablation}, \app outperforms three \gptleak variants (including two CoT variants) in overall detection effectiveness, achieving a higher F1 score with comparable BDR and much lower FAR. In particular, the two CoT variants outperform the straightforward \gptleak, indicating that the suitable design of CoT can help LLMs better identify bugs. In addition, \app  further outperforms the two CoT variants, indicating the benefits of combining LLMs and static analysis, which is also the main insight of \app. More detailed analyses are presented below.}

\revise{\textbf{Comparison among CoT variants and the basic \gptleak.}
Compared to the straightforward \gptleak, the two CoT strategies in \gptleakexp and \gptleakroi show  improvements in reducing false alarms while slightly decreasing the bug detection rate. For example, \gptleakroi reduces more than half of the false alarms compared to \gptleak by employing CoT instructions to first inferring resource-oriented intentions (i.e., the  same as \app). It actually confirms that our proposed intention inference is also beneficial for purely LLM-based resource leak detection. Overall, incorporating suitable CoT prompt is beneficial when applying LLMs for resource leak detection.}

\revise{
\textbf{Comparison between \app and CoT variants.}
Although the two CoT variants (i.e., \gptleakexp and \gptleakroi) outperform the basic \gptleak, \app further outperforms the two CoT variants in overall detection effectiveness.  In particular,  the static analysis can equip LLMs with more rigorous analysis, which can help reduce false positives in end-to-end LLM-based detection; while LLMs are also complementary to the static analysis as LLMs exhibit more general capabilities in code comprehension. As a result, by combining static analysis and LLMs, \app can achieve a balanced and effective resource leak detection mechanism, ultimately enhancing its overall accuracy and reliability.}

% An interesting observation is that \gptleakcot performs worse than \gptleak when utilizing the CoT instructions. After observing the results, we found that this issue arises because the CoT application causes the output format to be unstable, which prevents our regular expression from accurately extracting the answers.

% This phenomenon may suggest that the LLMs detect resource leaks using different logic than the procedure of static analysis-based detection. Explicitly inferring the three types of resource-oriented intentions proves to be beneficial for boosting static resource leak detection, but appears to be unnecessary for LLMs. This observation emphasizes the significance of prompt design when applying LLMs for resource leak detection.

% \revise{
% The results in Table~\ref{tab:ablation} also show certain impacts of the employed CoT prompts. Compared to \appempty, \app detects four more bugs but also reports one more false positive, resulting in a 5.0\% improvement in $F1$, indicating the usefulness of leading instructions in detecting more bugs. This also emphasizes the significance of CoT prompt design when applying LLMs for intention inference. Compared to \appexp, \app detects two more bugs and reports the same number of false positives, with a slight improvement of 0.7\% in $F1$. Given the occasional variability of the GPT-4 API's responses, we can conclude that the BDRs and FARs of \app and \appexp are generally consistent. This demonstrates the robustness of \app in utilizing different reasonable leading instructions.
% }

\begin{table}
	\centering
    \setlength{\tabcolsep}{8pt}
	\caption{\revise{Ablation Study Results}}\label{tab:ablation}
    \vspace{-2mm}
    \renewcommand{\arraystretch}{1.2}
    \begin{tabular}{l|rr|rr|r}
        \Xhline{2\arrayrulewidth}
        \textbf{Detector} & \textbf{\#TP} & \textbf{\#FP} & \textbf{BDR}~$\uparrow$ & \textbf{FAR}~$\downarrow$ & \textbf{F1}~$\uparrow$ \\ 
        \hline 
        \gptleak          & 57            & 58            & 66.3\%         & 67.4\%          & 0.567          \\
        \gptleakexp       & 52            & 37            & 60.5\%         & 43.0\%          & 0.594          \\
        \gptleakroi       & 49            & 26            & 57.0\%         & 30.2\%          & 0.609          \\
        \hline
        % \appempty         & 47            & 15            & 54.7\%         & 17.4\%          & 0.635          \\
        % \appexp           & 49            & 16            & 57.0\%         & 18.6\%          & 0.662          \\
        % \hline
        \app              & 51            & 16            & 59.3\%         & 18.6\%          & 0.667          \\ 
        \Xhline{2\arrayrulewidth}
    \end{tabular}
	% \end{adjustbox}
\end{table}

\finding{
    The results of the ablation study confirm the contributions of both LLM-based resource-oriented intention inference and static resource leak detection in \app. 
    % They also highlight the effectiveness of the leading instructions in the employed CoT prompts.
    % The effectiveness of \app in detecting resource leaks can be attributed to the synergistic utilization of these two key components, which enhances the overall detection performance.
}
\subsection{Generalizability with Different LLMs (RQ4)}
\revise{
We conduct an experiment to investigate the generalizability of \app when integrated with different LLMs.
}

\subsubsection{Selection of LLMs}
\revise{
In addition to the advanced closed-source \gptfour, we explore the effectiveness and efficiency of \app when integrated with the following two state-of-the-art open-source LLMs. 
\begin{itemize}[leftmargin=10pt]
    \item \textbf{Llama3-8B}~\cite{Llama3}: Meta's Llama 3 is a family of large language models available in various sizes, featuring both pretrained and instruction-tuned generative text models. We use Llama3-8B, which has 8 billion parameters.
    \item \textbf{Gemma2-9B}~\cite{Gemma2}: Google's Gemma 2 models are lightweight, instruction-tuned large language models, built using the same technology as the Gemini models~\cite{Gemini}. We employ the Gemma2-9B version, which has 9 billion parameters.
\end{itemize}}
%\revise{\noindent In our experiment, we use  \textit{Meta-Llama-3-8B}~\cite{Llama3-HF} and \textit{gemma-2-9b-it}~\cite{Gemma2-HF} versions available on the Hugging Face Platform, and run the model inference with the Transformers library~\cite{Transformers}. To align with the temperature setting (i.e., temperature=0) of \gptfour, we use greedy search decoding for both models, ensuring there is no randomness. The experiments are run on a workstation with 4 AMD EPYC 9554 64-Core CPUs, 512 GB of memory, and a single Nvidia H100 (80GB) GPU.} \yiling{I move them to implementation.}

\subsubsection{Results}
\revise{
Table~\ref{tab:generalizability} presents the comparison in the effectiveness and efficiency of \app when integrated with different LLMs. 
Overall, INFERROI with different LLMs consistently outperforms existing baselines (i.e., SpotBugs, Infer, and PMD)  in F1 score.
The results show that open-source Gemma2-9B achieves comparable effectiveness (0.644 vs. 0.667 in $F1$ score) with \gptfour, while Llama3-8B shows relatively lower metrics. 
% Comparing \gptfour and GPT-3.5, \gptfour demonstrates slight improvements in effectiveness metrics such as BDR, FAR, and F1, but requires a longer average response time. 
The two smaller open-source models, Llama3-8B and Gemma2-9B, have longer response times than the larger \gptfour due to the lack of optimizations like batch processing. 
%In practical usage, we can select the most suitable LLMs based on a trade-off between effectiveness, efficiency, stability, and cost.
% Note that the results returned by GPT-series LLMs are occasionally unstable in a few cases, even with a temperature setting of 0, while the results from the locally deployed models remain consistent.
}

\finding{
    \app demonstrates substantial generalizability when integrated with different LLMs, including closed-source \gptfour and open-source LLMs, maintaining relatively consistent effectiveness in resource leak detection.
}

\begin{table}
	\centering
    \setlength{\tabcolsep}{6pt}
	\caption{\revise{Results of Different LLMs on DroidLeaks Dataset}}\label{tab:generalizability}
    \vspace{-2mm}
    \renewcommand{\arraystretch}{1.2}
    \begin{tabular}{l|rr|rr|r|r}
        \Xhline{2\arrayrulewidth}
        \textbf{LLM} & \textbf{\#TP} & \textbf{\#FP}   & \textbf{BDR}~$\uparrow$  & \textbf{FAR}~$\downarrow$ & \textbf{F1}~$\uparrow$ & \textbf{Time} \\ 
        \hline
        Llama3-8B         & 39            & 18              & 45.3\%        & 20.9\%       & 0.545   & 7.1s   \\
        Gemma2-9B         & 48            & 15              & 55.8\%        & 17.4\%       & 0.644   & 11.4s  \\
        \hline 
        \gptfour          & 51            & 16              & 59.3\%        & 18.6\%       & 0.667   & 8.7s   \\
        % GPT-3.5           & 50            & 18              & 58.1\%        & 20.9\%       & 0.649   & 2.3s   \\
        \Xhline{2\arrayrulewidth}
    \end{tabular}
    % \begin{tabular}{l|rr|rr|r|r|r}
    %     \toprule
    %     \textbf{LLM} & \textbf{\#TP} & \textbf{\#FP}   & \textbf{BDR}~$\uparrow$  & \textbf{FAR}~$\downarrow$ & \textbf{F1}~$\uparrow$ & \textbf{Time} & \textbf{Cost (\$)} \\ 
    %     \midrule 
    %     \gptfour             & 51            & 16              & 59.3\%        & 18.6\%       & 0.667   & 8.7s   & 0.0128   \\
    %     GPT-3.5           & 50            & 18              & 58.1\%        & 20.9\%       & 0.649   & 2.3s   & 0.0004   \\
    %     \midrule
    %     Llama3-8B         & 39            & 18              & 45.3\%        & 20.9\%       & 0.545   & 7.1s   & ---      \\
    %     Gemma2-9B         & 48            & 15              & 55.8\%        & 17.4\%       & 0.644   & 11.4s  & ---      \\
    %     \bottomrule
    % \end{tabular}
    % \begin{tabular}{l|cc|cc|cc|c}
    %     \toprule
    %     \textbf{LLM} & \textbf{\#TP} & \textbf{\#FP}   & \textbf{BDR}  & \textbf{FAR} & \textbf{Precision} & \textbf{Recall} & \textbf{F1} \\ 
    %     \midrule 
    %     \gptfour             & 51            & 16              & 59.3\%        & 18.6\%       & 0.761    & 0.593     & 0.667         \\
    %     GPT-3.5           & 50            & 18              & 58.1\%        & 20.9\%       & 0.735    & 0.581     & 0.649         \\
    %     \midrule
    %     Llama3-8B         & 39            & 18              & 45.3\%        & 20.9\%       & 0.684    & 0.453     & 0.545       \\
    %     Gemma2-9B         & 48            & 15              & 55.8\%        & 17.4\%       & 0.762    & 0.558     & 0.644       \\
    %     \bottomrule
    % \end{tabular}
	% \end{adjustbox}
\end{table}

\subsection{Accuracy in Resource-Oriented Intention Inference (RQ5)}
In this evaluation, we evaluate the effectiveness of \app's LLM-based resource-oriented intention inference.

\subsubsection{Procedure}
To evaluate the effectiveness of \app's intention inference, two authors independently annotate all three types of resource-oriented intentions present in the 172 code snippets collected from the DroidLeaks dataset. In cases of disagreement, they engage in discussions to reach a final decision, ensuring consistent and accurate annotations. These manual annotations serve as the ground truth intentions ($GTs$) for the evaluation.
Subsequently, we apply \app to the code snippets to infer the intentions (denoted as $Preds$), and then calculate the \textit{precision} and \textit{recall} of the inference based on the manual annotations as follows:
% Precision represents the proportion of intentions inferred by \app that align with the ground truth, while recall indicates the proportion of true intentions from the ground truth that \app correctly infers. The precision and recall calculations are as follows:
$$\textit{precision} = \frac{|GTs \cap Preds|}{|Preds|};  \textit{recall} = \frac{|GTs \cap Preds|}{|GTs|}$$
% $$\textit{precision} = \frac{|Ground \cap Inferred|}{|Inferred|} $$
% $$\textit{recall} = \frac{|Ground \cap Inferred|}{|Ground|} $$

\revise{We compare the intention inference of \app with MiROK~\cite{wang2023mining}, which is a state-of-the-art technique for mining \revise{RAR (Resource Acquisition and Release)} API pairs. Specifically, we use the abstract RARs reported by MiROK to match APIs in the code for identifying intentions and then calculate precision and recall using the aforementioned equations. Note that MiROK does not support VAL intentions, so its metrics are calculated only for ACQ and REL intentions.} Furthermore, we assess the coverage of \app-inferred intentions for the 28 resource types in DroidLeaks and the 368 resource types in JLeaks, and then compare this coverage with that of PMD.

\subsubsection{Results}\label{sec:rq1-results}
\app achieves a precision of 74.6\% and a recall of 81.8\% in resource-oriented intention inference, indicating its ability to relatively accurately and comprehensively identify intentions within the code snippets. \revise{In contrast, intention identification based on MiROK yields only 54.0\% precision and 42.3\% recall, which is significantly lower than \app.}
Moreover, \app demonstrates superior versatility in addressing a broader spectrum of resource types, attaining a coverage of 67.9\% (19 types) for DroidLeaks and 60.1\% (221 types) for JLeaks, compared to PMD's coverage of 32.1\% (9 types) and 12.8\% (47 types), respectively.

Through analysis of the evaluation results, we attribute the effectiveness of \app's intention inference to the powerful capabilities of LLMs in leveraging background knowledge and understanding code context. For example, \app can identify the \inlinecode{AndroidHttpClient}-related intentions, which are ignored by other baseline detectors.

% Secondly, our well-designed prompt template plays a critical role in appropriately eliciting the relevant capabilities of LLMs. When we experimented with a prompt template lacking the first two leading instructions (see Section~\ref{sec:prompt}), the inference precision and recall experienced a decline of 20-30\%. This underscores the importance of prompt engineering for LLMs, e.g., guiding them with clear instruction-chains to enhance their reasoning abilities in intention inference. 

\finding{
    \app exhibits a precision of 74.6\% and a recall of 81.8\% in inferring resource-oriented intentions, achieving resource type coverages of 67.9\% and 60.1\% for DroidLeaks and JLeaks, respectively. This achievement can be attributed to the capabilities of LLMs, coupled with the efficacy of our designed prompt template.
    % These results underscore the effectiveness of \app in accurately and comprehensively identifying intentions within code.
}

\section{Discussion}

\subsection{Limitations}

% \subsection{Static Analysis}
% \app adopts a lightweight intraprocedural static analysis for detecting resource leaks. Instead of utilizing interprocedural analysis, we opt for an intraprocedural analysis based on the observation that many lower-level resource APIs (e.g., \inlinecode{AndroidHttpClient}) are encapsulated within higher-level resource APIs (e.g., \inlinecode{Socket}). By directly inferring intentions for higher-level resources, we can reduce the complexity of interprocedural program analysis during the detection process. 

\begin{figure}
    \centering
    \includegraphics[width=0.85\linewidth]{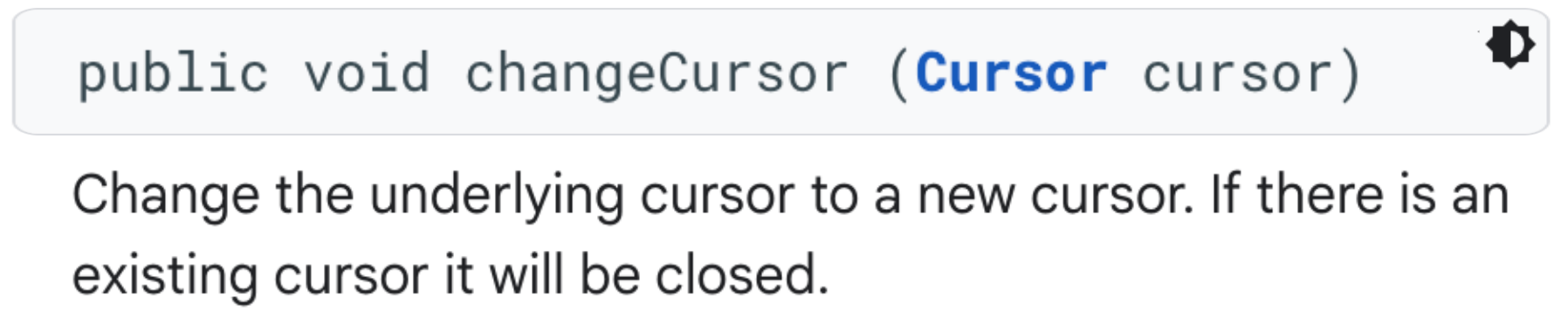}
    \vspace{-3mm}
    \caption{\revise{API Description of \inlinecode{CursorAdapter.changeCursor(Cursor)}}}
    \label{fig:api-doc}
\end{figure}

\revise{\textbf{Intention Inference.} 
Some reported false alarms arise because \app fails to recognize certain resource release intentions. For example, five false alarms in DroidLeaks result from \app not inferring that the \inlinecode{Cursor} resource is released when calling the Android method \inlinecode{CursorAdapter.changeCursor(Cursor)}. This oversight may be due to the API name \inlinecode{changeCursor} not explicitly indicating its resource release functionality, causing \app to overlook it. In fact, as shown in Figure~\ref{fig:api-doc}, the Android documentation specifies that \inlinecode{changeCursor(Cursor)} releases the existing \inlinecode{Cursor} resource. However, it appears that the LLM does not retain this information.
To address this issue, future work can explore integrating retrieval-augmented generation (RAG) techniques to search relevant API documentation and achieve more accurate and comprehensive intention inference.
}

\textbf{Static Analysis.}
It is important to acknowledge the limitations in the employed static analysis, especially in scenarios where certain types of resource leaks might remain undetected. For instance, resource leaks occurring within Android callbacks represent a challenging area where our lightweight static analysis may fall short.
Our key insights lie in directly inferring resource-oriented intentions based on code context understanding to boost static leak detection. The advantage of inferring resource-oriented intentions is its potential applicability to various static leak detection techniques, including callback-aware analyses like Relda2~\cite{wu2016relda2}. By integrating the inferred resource-oriented intentions into such existing techniques, we can enhance their capability to handle more comprehensive cases. In this regard, our LLM-based resource-oriented intention inference can provide \textit{complementary insights and improvement} for existing static resource leak detection, with the potential to empower a wide range of analysis methods to achieve more comprehensive and accurate results.

\subsection{Threats to Validity}
The internal validity of our studies is potentially affected by the randomness in data sampling and the subjectiveness in data annotation. To mitigate these threats, we adopted commonly-used data sampling strategies and involved multiple annotators to minimize any preference bias. 
\revise{Another concern is the risk of data leakage, as the evaluation data might have been incorporated into the training data of the utilized LLMs. To the best of our knowledge, there is no existing resource leak benchmark that is guaranteed to be exempted from data leakage issues (i.e., only containing data after \gptfour training data cutoff). Therefore, to mitigate this issue, we have made the following attempts to justify that the improvements of \app are not simply caused by the potential data memorization. (i) We evaluate InferROI to find 26 previously-unknown bugs when scanning the open-source projects in RQ2. Even if the code itself could be seen by LLMs during its training, the previously-unknown bugs are not seen by LLMs before. (ii) We show \app outperforms basic \gptfour in the ablation study (RQ3.b). It indicates that the improvements of \app do not simply come from the potential memorization of \gptfour on similar examples from training. (iii) We include  multiple datasets in RQ1 to address the overfitting issues. In addition to the classic benchmark DroidLeak, we also include the latest resource leak benchmark JLeaks which has shown to have better diversity and quality than previous benchmarks.% Jleaks paper was released just one week before the submission time, and we tried our best to include it to mitigate the overfitting issues by including more diverse testing samples.
}
% However, the ablation study (RQ3.b) already shows that purely querying \gptfour itself to detect resource leak has poor effectiveness, indicating that the effectiveness of \app is mainly credited to our proposed intention inference and static analysis algorithm not to the data leakage. Furthermore, we further curate projects created after the date of the \gptfour's training data collection (in RQ2) to mitigate this issue.
As for the external validity, the evaluation data used in our work  might not guarantee the full generality of our findings. To address this concern, we evaluate \app on both existing datasets and real-world open-source projects to detect resource leaks. 
% Nonetheless, exploring the extension of \app to other programming languages and integrating \app with more analysis techniques could be intriguing avenues for future research. 
\section{Related Work}
\textbf{Resource Leak Detection.}
Automated resource leak detection techniques~\cite{torlak2010effective, wu2016light, kellogg2021lightweight, liu2019droidleaks} have been proposed to detect whether some resource is not being released after its acquisition. Typically there are two important components for resource leak detection. First, identify the potential API pairs for resource acquisition and resource release; then based on the RAR pairs, analyze the code to check whether the release API is not subsequently called after the acquisition API. 
The majority of existing resource leak detection techniques are concentrated on the analysis part by proposing more precise and more scalable code analysis approaches~\cite{torlak2010effective, wu2016light, kellogg2021lightweight, emamdoost21ndss, saha2013hector, li2020pca}. For example, Torlak et al.~\cite{torlak2010effective} combine intra-procedural analysis and inter-procedural analysis to enable more scalable and more accurate detection for system resource leaks (e.g., I/O stream and database connections); Wu et al.~\cite{wu2016light} propose an inter-procedural and callback-aware static analysis approach to detect resource leak in Android apps; Kellogg et al.~\cite{kellogg2021lightweight} incorporate ownership transfer analysis, resource alias analysis, and fresh obligation creation to enable more precise analysis. 
The API pairs used in most existing techniques~\cite{torlak2010effective, wu2016light} are often predefined by human expertise and heuristic rules, which not only require non-trivial human efforts but also have limited coverage in libraries and APIs. For example, Torlak et al.~\cite{torlak2010effective} manually collect API pairs and detect resource leaks for \textit{stream} and \textit{database} in JDK. In addition, SpotBugs~\cite{spotbugs} only considers the predefined \textit{stream} related API pairs and thus could only detect \textit{stream} related resource leaks. 
More recently, Bian et al.~\cite{bian2020sinkfinder} propose SinkFinder, which mines and classifies frequent resource-related API pairs for a given project.

Our work does not rely on prepared API pairs and directly infers resource-oriented intentions in code by utilizing the exceptional code comprehension capability of LLMs.

%Our evaluation results in Section~\ref{sec:rq2} also indicate that our approach substantially outperforms SinkFinder by generating much more high-quality RAR pairs when applied to a large number of libraries. 

\textbf{LLMs for Software Engineering.}
% In other domains, there are also some techniques that  mine API usage patterns to detect the relevant API misuse~\cite{li2005pr, acharya2007mining, chang2007finding, zhong2009mapo, nguyen2009graph, nguyen2014mining, yun2016apisan, zhang2018code, bian2020sinkfinder, nielebock2021guided}. For example, Chang et al.~\cite{chang2007finding} leverage frequent subgraph and itemset mining to mine rules to detect neglected conditions. ExampleCheck~\cite{zhang2018code} mines 180 API usage patterns for 100 popular Java APIs to detect API misuses such as missing control constructs and incorrect guard conditions. Our work targets a domain (i.e., resource leak) different from these techniques, i.e.,  we focus on mining the usage patterns of APIs on resource-operation knowledge. To this end, we not only propose a novel representation (i.e., absRAR pairs) to represent the resource-knowledge-related API usage pairs, but also propose a novel learning-based mining approach to mine such absRAR pairs from a large code corpus. 
LLMs have been widely used in software engineering tasks~\cite{chen2024deep,wang2024teaching,yuan2024evaluating,du2024evaluating,du2024vul,liu2024large,zhang2024vuladvisor,zhang2024empirical,hou2023large,zheng2023survey,wang2024agents,wang2024rlcoder}.
Recent studies have demonstrated that pre-trained language models can be utilized as knowledge bases by using synthetic tasks similar to the pre-training objective to retrieve the knowledge/information stored in the models~\cite{petroni19llmkb,jiang2020can}. These works have shown that language models can recall factual knowledge without any fine-tuning by using proper prompts.
Several works have focused on exploring effective prompt templates to improve performance of PLMs on downstream problems. In a recent survey by Liu et al.~\cite{liu23promptsurvey}, prompt templates are broadly classified into two categories: \textit{cloze prompts}~\cite{cur21templatener,petroni19llmkb,wang2024tiger}, which entail filling in the blanks in text or code, and \textit{prefix prompts}~\cite{lester21prompt,li20prefixtuning}, which continue generating content following a specified prefix. 
Recent studies have explored the application of prompt learning in various software engineering tasks. Wang et al.~\cite{wang22experimental} investigated the effectiveness of prompt learning in code intelligence tasks such as clone detection and code summarization. Huang et al.~\cite{huang22typeinfer} used prompt learning for type inference. 

This works consider LLMs as knowledge bases of resource management and prompts them to infer resource-oriented intentions based on their powerful code understanding capability.
\section{Conclusions and Future Work}
In this work, we propose \app, a novel resource leak detection approach which boosts static analysis via LLM-based resource-oriented intention inference. Given a code snippet, \app prompts the LLM in inferring involved intentions from the code. By aggregating these inferred intentions, \app utilizes a lightweight static-analysis based algorithm to analyze control-flow paths extracted from the code, thereby detecting resource leaks. We evaluate \app for the Java programming language and investigate its effectiveness in resource leak detection. 
Experimental results on the DroidLeaks and JLeaks datasets demonstrate that \app exhibits promising bug detection rate (about 60\%) and false alarm rate (about 20\%). Moreover, when applied to open-source projects, \app identifies 29 unknown resource leak bugs. In the future, we will explore the extension of \app to other programming languages and integrate \app with more program analysis techniques. 

\section*{Data Availability}
To facilitate further research, we have released our code and data in the replication package~\cite{replication}.

\section*{Acknowledgement}
This research project is supported by National Key R\&D Program of China (2023YFB4503805) and National Natural Science Foundation of China under Grant No. 62302099.

\clearpage

\bibliographystyle{IEEEtran}
\balance
\bibliography{llm4leak-main.bib}

\end{document}